\documentclass{emulateapj}
\usepackage{apjfonts}
\usepackage{epsfig}
\usepackage{amsmath}
\usepackage{amssymb}
\usepackage{amsfonts}
\usepackage{lscape}
\usepackage{natbib}          
\shorttitle{VARIABILITY IN THE LMXBS OF NGC~4697}
\shortauthors{SIVAKOFF ET AL.}

\defcitealias{JCF+2004}{J2004}
\defcitealias{BIS+2006}{B2006}
\defcitealias{SJS+2007}{S2007}

\defcitealias{ISB2000}{I2000}
\defcitealias{SIB2000}{Paper~I}
\defcitealias{SIB2001}{Paper~II}
\defcitealias{SSJ2005}{Paper~III}
\defcitealias{SJJ+2008a}{Paper~IV}
\defcitealias{SJJ+2008b}{Paper~V} 

\newcommand\cennodata{\multicolumn{1}{c}{\nodata}}

\begin{document}

\title{Measurements of Variability of Low Mass X-ray Binary Candidates
in the Early-Type Galaxy NGC~4697 from Multi-Epoch Chandra X-ray Observations}
\author{
Gregory R. Sivakoff\altaffilmark{1,2},
Andr\'{e}s Jord\'{a}n\altaffilmark{3,4},
Adrienne M. Juett\altaffilmark{5},
Craig L. Sarazin\altaffilmark{1},
Jimmy A. Irwin\altaffilmark{6}
}

\altaffiltext{1}{
Department of Astronomy,
University of Virginia,
P. O. Box 400325,
Charlottesville, VA 22904-4325, USA;
sarazin@virginia.edu}
\altaffiltext{2}{
Current Address:
Department of Astronomy,
The Ohio State University,
4055 McPherson Laboratory
140 W. 18th Avenue, Columbus, OH 43210-1173, USA;
sivakoff@astronomy.ohio-state.edu}
\altaffiltext{3}{%
Clay Fellow,
Harvard-Smithsonian Center for Astrophysics,
60 Garden Street,
MS-67, Cambridge, MA 02138, USA;
ajordan@cfa.harvard.edu}
\altaffiltext{4}{%
Departamento de Astronom\'{\i}a y Astrof\'{\i}sica, 
Pontificia Universidad Cat\'olica de Chile,
Casilla 306, Santiago 22, Chile}
\altaffiltext{5}{%
NASA Postdoctoral Fellow,
Laboratory for X-ray Astrophysics,
NASA Goddard Space Flight Center,
Greenbelt, MD 20771, USA;
adrienne.m.juett@nasa.gov}
\altaffiltext{6}{
Department of Astronomy,
909 Dennison Building,
University of Michigan,
Ann Arbor, MI 48109-1042, USA;
jairwin@umich.edu}

\begin{abstract}
Multi-epoch {\it Chandra} X-ray observations of nearby massive early-type
galaxies open up the study of an important regime of low-mass X-ray binary
(LMXB) behavior --- long term variability. In a companion paper, we report on
the detection of 158 X-ray sources down to a detection/completeness limit of
$0.6/1.4\times 10^{37} {\rm \, ergs \, s}^{-1}$ using five {\it Chandra}
observations of NGC~4697, one of the nearest ($11.3 {\rm \, Mpc}$), optically
luminous ($M_B < -20$), elliptical (E6) galaxy. In this paper, we report on the
variability of LMXB candidates measured on timescales from seconds to years. At
timescales of seconds to hours, we detect five sources with significant
variability. Approximately $7\%$ of sources show variability between any two
observations, and $16\pm4\%$ of sources do not have a constant luminosity over
all five observations. Among variable sources, we identify eleven transient
candidates, with which we estimate that if all LMXBs in NGC~4697 are long-term
transients then they are on for $\sim 100 {\rm \, yr}$ and have a 7\% duty
cycle. These numbers are consistent with those found for brighter LMXBs in M87
and NGC~1399, which suggests that there does not appear to be a measurable
difference between the outburst durations of long-term transient neutron star
LMXBs and black hole LMXBs. We discuss in detail a transient supersoft source,
whose properties are not easily explained by standard explanations for supersoft
sources.
\end{abstract}

\section{Introduction}
\label{sec:n4697x_intro}

With the angular resolution of the \textit{Chandra X-Ray Observatory} low-mass
X-ray binaries (LMXBs) in nearby early-type galaxies can be detected. For
early-type galaxies with a relatively low X-ray--to--optical luminosity ratio,
known as X-ray faint galaxies, the emission from LMXBs is the dominant source
\citep[e.g.,][hereafter Papers I and II]{SIB2000,SIB2001}.
In typical {\it Chandra} observations of such galaxies, tens to a few hundred
LMXBs are resolved at bright X-ray luminosities, $L_X \gtrsim 5$--$10\times
10^{37} {\rm \, ergs \, s}^{-1}$. By stacking multi-epoch {\it Chandra}
observations of early-type galaxies, larger numbers of fainter LMXB candidates
can be studied, and brighter LMXB candidates can be studied in greater detail. 
Such observations open up the study of an important regime of behavior --- long
term variability. Low mass X-ray binaries exhibit a wide range of variability,
including transient outbursts, type I bursts, and milder fluctuations. Our
understanding of LMXB variability in the Galaxy drives models for accretion and
LMXB evolution. However, there are several limitations in studying Galactic
LMXBs: source distances are known for only a small subset, it is difficult to
observe the whole Galaxy at once, and the size of the observed sample is very
limited. Observations of nearby early-type galaxies can overcome these
limitations of Galactic studies.

Examples of extragalactic LMXB candidates with extreme behaviors, such as
luminous supersoft sources (SSs) and flaring sources, have also been examined
with {\it Chandra} in greater detail. In our Galaxy and M31, SSs have very soft
X-ray spectra $(\lesssim 75 {\rm \, eV})$ and are generally believed to be
accreting white dwarfs (WDs). Since the bolometric luminosities of extragalactic
SSs can exceed the Eddington luminosity for a Chandrasekhar mass WD, an
alternate hypothesis that these sources contain intermediate-mass ($\sim
10^2$--$10^3 \, M_{\sun}$) accreting BHs has been put forth \citep{SGS+2002}. 
Due to the degradation of the soft response of the Advanced CCD Imaging
Spectrometer (ACIS) on {\it Chandra}, the study of SSs has been limited. Some
LMXBs exhibit another extreme behavior, namely relatively short timescale
(seconds to minutes) flares. Recently, {\it XMM-Newton} detected type I X-ray
bursts in M31 \citep{PH2005} and we reported a more extreme example of flares in
NGC~4697 \citetext{\citealp{SSJ2005}, hereafter \citetalias{SSJ2005}}.

In this paper, we report on the variability of LMXB candidates measured from
multi-epoch {\it Chandra} X-ray observations of NGC~4697, one of the nearest
($11.3 {\rm \, Mpc}$), optically luminous ($M_B < -20$), elliptical (E6) galaxy. 
In a companion paper (Sivakoff et al 2008b, hereafter Paper IV) we
report on the properties of LMXB candidates excluding variability, as well as
the connection to globular clusters (GCs) revealed by a {\it Hubble Space
Telescope} Advance Camera for Surveys ({\it HST}-ACS) observation near the
galaxy center.

In \S~\ref{sec:n4697x_obs}, we recap key aspects of the observations and data
reduction of NGC~4697 discussed in Paper IV. Our observations of NGC~4697
provide five approximately equal chances to detect variability within a given
$\sim 40 {\rm \, ks}$ observation (\S~\ref{sec:n4697x_src_var_intra}), as well
as variability between different observations on timescales of $11 {\rm \, d}$
to $4.6 {\rm \, yr}$ (\S~\ref{sec:n4697x_src_var_inter}). We summarize our
conclusions in \S~\ref{sec:n4697x_conclusion}. Unless otherwise noted, all
errors refer to $1 \sigma $ confidence intervals, count rates are in the
$0.3$--$6 {\rm \, keV}$ band, and fluxes and luminosities are in the $0.3$--$10
{\rm \, keV}$ band.

\section{Chandra Observations and Data Reduction}
\label{sec:n4697x_obs}

Since detailed descriptions of the {\it Chandra} observation and
data reduction are reported in Paper IV, we only recap key aspects here.
This analysis covers five {\it Chandra} ACIS observations
(Observations
\dataset[ADS/Sa.CXO#obs/00784]{0784},
\dataset[ADS/Sa.CXO#obs/04727]{4727},
\dataset[ADS/Sa.CXO#obs/04728]{4728},
\dataset[ADS/Sa.CXO#obs/04729]{4729}, and
\dataset[ADS/Sa.CXO#obs/04730]{4730}) taken on 
2000 January 15, 2003 December 26, 2004 January 06, February 02, and August 18. 
The final flare-filtered live exposure times for the five observations were
37174, 39919, 35601, 32038, and $40044 {\rm \, s}$. There are differences in the
detector setup and instrument properties, such that the response of the Cycle-1
observation (0784) differs most from the Cycle-5 observations (4727, 4728, 4729,
4730). Key differences include the gain files used
(acisD1999-09-16gainN0005.fits for Observation 0784 versus
acisD2000-01-29gain\_ctiN0001.fits with time-dependent gain and charge transfer
inefficiency corrections for the Cycle-5 observations), smaller background due
to Very-Faint mode cleaning in Cycle-5 observations, and increasing quantum
efficiency (QE) degradation with time.

All {\it Chandra} observations were analyzed using {\sc ciao 3.1}%
\footnote{See \url{http://asc.harvard.edu/ciao/}.} 
with {\sc caldb 2.28} and NASA's {\sc ftools 5.3}%
\footnote{See
\url{http://heasarc.gsfc.nasa.gov/docs/software/lheasoft/}%
\label{ftn:heasoft}.}. Source positions and extraction regions were
refined using ACIS Extract 3.34%
\footnote{See \url{http://www.astro.psu.edu/xray/docs/TARA/%
ae\_users\_guide.html}.}. All spectra
were fit using {\sc xspec}\footnotemark[\ref{ftn:heasoft}].

The finding charts for the sources are displayed in Figures 1 and 2 of Paper IV. 
In that Paper, the source properties from stacking the multi-epoch observations
are listed in Tables 1 and 7, and the instantaneous luminosities for each
observation are given in Table 6.

Throughout Paper IV and this paper, we refer to the Analysis Sample, which
includes 126 sources that have photometric count rates determined at the $\ge 3
\sigma$ level when stacking all five observations. This corresponds to having
at least 18 net counts, a minimum detected count rate of $1.0 \times 10^{-4}
{\rm \, counts \, s}^{-1}$, and a minimum luminosity of $1.4 \times 10^{37} {\rm
\, ergs \, s}^{-1}$.

The spectral properties of sources can be crudely characterized by hardness
ratios or X-ray colors (e.g., \citetalias{SIB2000,SIB2001}). We defined hardness
ratios of $H_{21} \equiv (M - S)/(M + S)$, $H_{31} \equiv (H - S)/(H + S)$, and
$H_{32} \equiv (H - M)/(H + M)$, where $S$, $M$, and $H$ are the total counts in
the soft (0.3--1 keV), medium (1--2 keV), and hard (2--6 keV) bands
\citep{SSC2004}.

\section{Intraobservation Variability}
\label{sec:n4697x_src_var_intra}

After applying barycentric corrections to the times in event files, we
applied three techniques to search for intraobservation variability:
the Rayleigh statistic \citep[e.g.,][]{MBB+2003}, the K-S test
\citepalias[e.g.,][]{SIB2001}, and a newly developed flare detection technique
\citepalias{SSJ2005}.
These three techniques consider all photons in the source apertures,
and thus are likely to include some background photons.

\tabletypesize{\footnotesize}

\begin{deluxetable}{lccr}
\setlength{\tabcolsep}{0.05in}
\tablewidth{0pt}
\tablecaption{Possible Periodically Variable Sources
\label{tab:n4697x_pervar}}
\tablehead{
\colhead{}&
\colhead{}&
\colhead{}&
\colhead{Period $\tau_{\rm R}$}\\
\colhead{Source}&
\colhead{Observation}&
\colhead{Probability Constant $P_{\rm R}$}&
\colhead{(sec)}\\
\colhead{(1)}&
\colhead{(2)}&
\colhead{(3)}&
\colhead{(4)}
}
\startdata
\phn\phn1 & 0784 & 1.1E-02 &   21.93\\
101 & 4727 & 6.7E-03 &   14.98\\
103 & 4728 & 4.4E-02 &   20.02\\
143 & 0784 & 4.6E-08 & 1000.00\\
155 & 4728 & 8.1E-03 &   16.99\\
\enddata
\tablecomments{The value of $P_{\rm R}$ indicates the
probability that the observed periodicity is due to a statistical
fluctuation. All periods, $\tau_{\rm R}$, are measured in seconds.  No
periodic variable sources were detected in Observations 4729 and
4730. }
\setlength{\tabcolsep}{6pt}
\end{deluxetable}

We used the Rayleigh statistic to search for periodic signals with
frequencies between $5.0 \times 10^{-5} {\rm \, Hz}$ and $1.0 \times
10^{-1} {\rm \, Hz}$ (periods of $10$--$20,000 {\rm \, s}$), testing
every $2.5 \times 10^{-5} {\rm \, Hz}$. Table~\ref{tab:n4697x_pervar}
summarizes the results for $> 2
\sigma$ level variable sources (i.e., the chance the observed source
variability is due to a statistical fluctuation is $< 4.6$\%),
listing the observation, probability that the detected periodicity is due to a
statistical fluctuation, $P_{\rm R}$, and detected period, $\tau_{\rm
R}$.  Between zero and two sources appear periodic at the $> 2 \sigma$
level in any given observation. Since the expected number of false
detected sources at this level is $\sim 7$, we concentrate our
discussion on the one periodic source detected at the $> 3 \sigma$
level
(probability that the variability is due to a fluctuation is $< 0.27$\%).
The probability that the apparent periodicity among the 26
counts of Source 143 in Observation 0784 is due to a statistical
fluctuation is only $4.6\times10^{-8}$.
Unfortunately, this source is located near an edge of the S3 chip,
and the period of $\sim 1000 {\, \rm s}$ is consistent with a
periodicity induced by the yaw of the satellite aspect motion.
Thus, it is likely that this periodicity is instrumental, and we have no cases
of detected periodic variations
for sources in NGC~4697.

\begin{figure}
\plotone{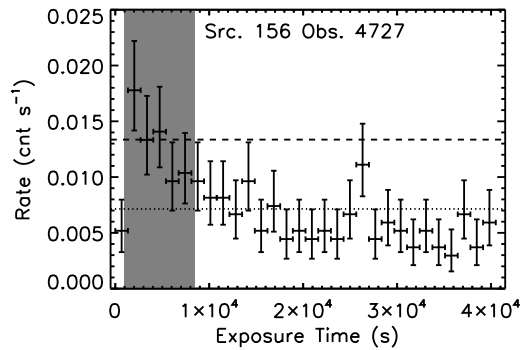}
\caption[X-ray Light Curve of Source 156 in NGC~4697]{
Binned lightcurve using 1350 ${\rm \, s}$ bins of Source 156
(the foreground star BD-05 3573) in Observation 4727.
Both the K-S test and our flare detection technique classify this source as
variable within this observation. This star flared shortly after the
observation began.
The average count rate during the observations is
indicated with the dotted line.
The count rate during the flare determined by our flare detection
algorithm is indicated by the dashed line and the extent of the flare is
indicated by the gray region.
\label{fig:n4697x_lng_flr}}
\end{figure}

\tabletypesize{\footnotesize}

\begin{deluxetable}{lcc}
\setlength{\tabcolsep}{6pt}
\tablewidth{0pt}
\tablecaption{KS Test for Intraobservation Variability
\label{tab:n4697x_ksvar}}
\tablehead{
\colhead{Source}&
\colhead{Observation}&
\colhead{$P_{\rm KS}$}\\
\colhead{(1)}&
\colhead{(2)}&
\colhead{(3)}
}
\startdata
\phn\phn2 & 0784 & 2.4E-02\\
\phn10 & 4727 & 4.6E-02\\
\phn11 & 0784 & 1.4E-02\\
\phn12 & 4728 & 1.2E-02\\
\phn40 & 0784 & 2.9E-02\\
\phn42 & 4730 & 3.0E-02\\
\phn54 & 4729 & 1.2E-02\\
\phn65 & 4727 & 1.6E-02\\
\phn71 & 4728 & 5.0E-03\\
\phn75 & 4727 & 2.0E-02\\
\phn76 & 4729 & 4.3E-02\\
\phn78 & 0784 & 1.5E-02\\
\phn79 & 0784 & 1.5E-02\\
\phn83 & 0784 & 2.6E-02\\
\phn96 & 0784 & 3.5E-02\\
100 & 4729 & 1.6E-02\\
115 & 4728 & 3.9E-02\\
130 & 0784 & 2.0E-02\\
137 & 0784 & 3.7E-02\\
    & 4727 & 2.0E-02\\
156 & 4727 & 6.9E-07\\
\enddata
\tablecomments{The value of $P_{\rm KS}$ indicates the probability that the observed
variability is due to a statistical fluctuation.}
\setlength{\tabcolsep}{6pt}
\end{deluxetable}

By comparing the cumulative fraction of events received from a source
to a constant rate, the K-S test can identify intraobservation variability.
This method has been applied to LMXB candidates in
previous papers \citepalias[e.g.,][]{SIB2001}. We summarize the
results in Table~\ref{tab:n4697x_ksvar} for sources which are variable
at the $> 2 \sigma$ level.
We list the
probability the detected variability is
due to a statistical fluctuation for each observation, $P_{\rm KS}$.
Between one and nine sources appear variable in any given observation.
Again, we concentrate on the one variable source detected at the $> 3 \sigma$
level.
Source 156 is the foreground star \object{BD-05 3573}
which undergoes a clear flare
($P_{\rm KS} = 6.9\times 10^{-7}$) soon after Observation 4727 begins.
We display the binned lightcurve of this source in
Figure~\ref{fig:n4697x_lng_flr}.
We note that source 137 appears as a $2 \sigma$
variable source in both
Observations 0784 and 4727.
Following equation 3 of
\citetalias{SSJ2005}, the joint probability that this source is constant
is $4.6\times 10^{-3}$, which is still less than $3 \sigma$ significant.

Finally, we applied a new method developed to search for flaring sources. This
technique is based on the arrival times of individual events compared to the
Poisson distribution at a constant rate, and was presented previously in
\citetalias{SSJ2005}. This flare detection uses the ONTIME values (37651, 40447,
36072, 32462, and $40574 {\rm \, s}$, respectively for observations 0784 and
4727--4730) as opposed to the live exposure times. \citetalias{SSJ2005} gave the
results for the most significantly flaring sources; note that Sources A, B, and
C in that Paper correspond to Sources 71, 75, and 57 here, respectively. We
summarize the results of these tests in Table~\ref{tab:n4697x_flarevar} for
sources which are variable at the $> 2 \sigma$ level in at least one
observation. For each observation of a flare-candidate source, we list the
observation number, the total number of photons in the observation, $N$, the
number of photons in the flare, $n$, the duration of the flare, $\delta t$, and
the Modified Julian Date (MJD) of the arrival time of the first photon in the
flare, $t_0$. The value of $P_{\rm constant}$ indicates the mathematical
probability of the given flare occurring due to a statistical fluctuation in a
constant rate source; however, we must account for the number of non-independent
flare searches undertaken. After applying corrections to $P_{\rm constant}$
based on the number of photons in the observation, $P_{\rm constant}^{\prime}$
is the probability a flare in a given observation is not real. The number of
sources with a statistically significant flare ranges from one to seven in the
different observations.

\begin{figure}
\begin{center}
\epsfig{file=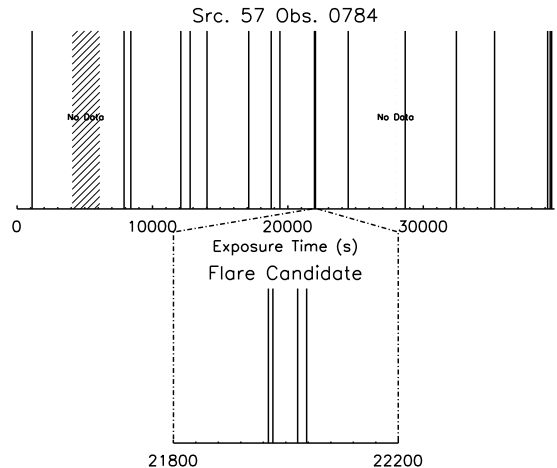, angle=90, width=0.85\linewidth,
       clip=, bbllx=0, bblly=65, bburx=540, bbury=700}
\end{center}
\caption[Photon Impulse Diagram of Source 57 in NGC~4697]{
Impulse diagram indicating the time of arrival of photons in 
Source 57 (Observation 0784). This LMXB candidate undergoes
short, bright flares in multiple observations
\citepalias{SSJ2005},
which were detected only with our new flare detection technique.
\label{fig:n4697x_sh_flr}}
\end{figure}

Given the number of sources expected to show a flare from a statistical
fluctuation, we only consider the flaring sources in individual observations
detected at the $> 3 \sigma$ level, which are Sources 57, 155, and 156. These
sources have flares ranging from a few photons over tens of seconds to nearly a
hundred photons over thousands of seconds.
Source 57 was discussed earlier in \citetalias{SSJ2005} as Source C; note that
the luminosity conversion used in this Paper is about 10\% smaller than in
\citetalias{SSJ2005} due to updated calibrations. We display an impulse diagram
of the photon arrival times for Observation 0784 of Source 57 in
Figure~\ref{fig:n4697x_sh_flr}. The observed count rate of the flare is over 100
times the average count rate. Source 57 has no optical counterpart in the
HST-ACS image, suggesting it is most likely a flaring Field-LMXB.

\tabletypesize{\footnotesize}

\begin{deluxetable*}{lcrrrrrrr}
\setlength{\tabcolsep}{6pt}
\tablewidth{0pt}
\tablecaption{Possible Flaring Sources in NGC~4697
\label{tab:n4697x_flarevar}}
\tablehead{
\colhead{Source}&
\colhead{Observation}&
\colhead{$N$} &
\colhead{$n$} &
\colhead{$\Delta t$} &
\colhead{$t_0$} &
\colhead{$P_{\rm constant}$}&
\colhead{$P_{\rm constant}^\prime$}&
\colhead{$P_{\rm constant, joint}$}\\
&
&
&
&
\colhead{(s)}&
\colhead{(MJD)}&
&
&
\\
\colhead{(1)}&
\colhead{(2)}&
\colhead{(3)}&
\colhead{(4)}&
\colhead{(5)}&
\colhead{(6)}&
\colhead{(7)}&
\colhead{(8)}&
\colhead{(9)}
}
\startdata
\phn25 & 4730             &       21 &        9 &     2952 & 53235.81643 &  2.5E-03 &  6.5E-03 & 3.2E-02\\
\phn42 & 4728             &       26 &        3 &       25 & 53010.59465 &  3.9E-03 &  1.0E-02 & 5.2E-02\\
\phn56 & 4727             &       10 &        5 &     1919 & 52999.89337 &  8.7E-03 &  2.1E-02 & 1.1E-01\\
\phn57 & 0784             &       20 &        4 &       68 & 51558.95770 &  1.3E-04 &  3.4E-04 &\cennodata\\
    & 4727             &       20 &        3 &       50 & 53000.06223 &  5.5E-03 &  1.4E-02 &\cennodata\\
    & 0784, 4727       &\cennodata&\cennodata&\cennodata& \cennodata  &\cennodata&\cennodata& 4.8E-05\\
\phn71 & 0784             &       14 &        5 &     1047 & 51558.79423 &  7.0E-03 &  1.8E-02 &\cennodata\\
    & 4727             &       16 &        4 &      628 & 52999.75882 &  2.7E-02 &  7.0E-02 &\cennodata\\
    & 4728             &       14 &        4 &      509 & 53010.70295 &  1.2E-02 &  3.1E-02 &\cennodata\\
    & 0784, 4727, 4728 &\cennodata&\cennodata&\cennodata& \cennodata  &\cennodata&\cennodata& 3.9E-04\\
\phn75 & 4727             &       10 &        5 &     1329 & 53000.04326 &  2.2E-03 &  5.5E-03 &\cennodata\\
    & 4728             &        9 &        4 &     1420 & 53010.72444 &  3.4E-02 &  8.0E-02 &\cennodata\\
    & 4729             &        6 &        5 &     5654 & 53047.59915 &  4.3E-02 &  9.0E-02 &\cennodata\\
    & 4727, 4728       &\cennodata&\cennodata&\cennodata& \cennodata  &\cennodata&\cennodata& 4.4E-03\\
    & 4727--4729       &\cennodata&\cennodata&\cennodata& \cennodata  &\cennodata&\cennodata& 4.0E-04\\
\phn78 & 0784             &       44 &        4 &       52 & 51558.84209 &  1.5E-03 &  4.1E-03 & 2.0E-02\\
\phn83 & 0784             &       10 &        8 &     6910 & 51558.83658 &  8.5E-03 &  2.1E-02 & 1.1E-01\\
\phn90 & 4727             &       13 &        3 &      116 & 52999.82502 &  7.5E-03 &  1.9E-02 & 9.5E-02\\
107 & 0784             &        2 &        2 &      480 & 51559.09218 &  2.5E-02 &  5.2E-02 &\cennodata\\
    & 4728             &        4 &        2 &       63 & 53010.69939 &  2.1E-02 &  4.2E-02 &\cennodata\\
    & 0784, 4728       &\cennodata&\cennodata&\cennodata& \cennodata  &\cennodata&\cennodata& 2.2E-02\\
115 & 4728             &       22 &        8 &     1857 & 53010.37696 &  2.7E-03 &  6.9E-03 & 3.5E-02\\
118 & 4728             &       35 &       11 &     2667 & 53010.42386 &  9.0E-03 &  2.5E-02 & 1.2E-02\\
124 & 4728             &       20 &        4 &      270 & 53010.38518 &  8.5E-03 &  2.2E-02 & 1.1E-01\\
135 & 0784             &        4 &        2 &       78 & 51558.70911 &  2.4E-02 &  5.0E-02 &\cennodata\\
    & 4730             &        6 &        2 &       57 & 53235.77593 &  4.1E-02 &  8.5E-02 &\cennodata\\
    & 0784, 4730       &\cennodata&\cennodata&\cennodata& \cennodata  &\cennodata&\cennodata& 4.3E-02\\
137 & 4727             &       27 &       12 &     5529 & 53000.01065 &  2.4E-02 &  6.5E-02 &\cennodata\\
    & 4729             &       18 &        6 &     1034 & 53047.40640 &  4.2E-03 &  1.1E-02 &\cennodata\\
    & 4727, 4729       &\cennodata&\cennodata&\cennodata& \cennodata  &\cennodata&\cennodata& 6.9E-03\\
146 & 0784             &       13 &        4 &      697 & 51559.01802 &  1.9E-02 &  4.9E-02 &\cennodata\\
    & 4727             &        4 &        2 &       22 & 52999.76129 &  6.5E-03 &  1.3E-02 &\cennodata\\
    & 0784, 4727       &\cennodata&\cennodata&\cennodata& \cennodata  &\cennodata&\cennodata& 6.6E-03\\
155 & 4728             &      219 &       17 &      584 & 53010.48733 &  2.2E-04 &  7.5E-04 & 3.7E-03\\
156 & 4727             &      289 &       99 &     7423 & 52999.64230 &  4.0E-06 &  1.3E-05 & 6.7E-05\\
\enddata
\tablecomments{This table follows the nomenclature of \citetalias{SSJ2005}:
$N$ is the total number of photons during an observation.
A flare candidate with $n$ photons of observed duration $\delta t$ begins at
$t_{0}$, where we list the time measured prior to barycentric corrections.
The value of
$P_{\rm constant}$ indicates the mathematical probability of the given
flare occurring due to a statistical fluctuation; $P_{\rm
constant}^{\prime}$ is the same probability after accounting for the
number of non-independent flare searches the technique examines. If the
same type of flare occurs in multiple observations, one can calculate
the joint probability the flare is not real, $P_{\rm constant, joint}$.}
\setlength{\tabcolsep}{6pt}
\end{deluxetable*}

\begin{figure}
\plotone{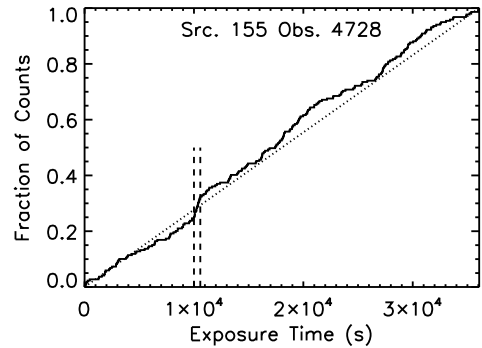}
\caption[Accumulated Fractional X-ray Light Curve of Source 155 in NGC~4697]
{
Cumulative lightcurve of Source 155 (Observation 4727) overlaid by the
expected distribution for a constant source (dotted line).  Our flare
detection technique classifies this source as variable within this
observation; however, the K-S test misses the variability of this source. 
The flare interval, which lies between the dashed lines, has a clearly increased
rate (larger slope).
\label{fig:n4697x_med_flr}}
\end{figure}

A flare of several hundred seconds with 17 counts is detected in Source 155, but
was missed by the K-S test. This source was just above the criteria used in
\citetalias{SSJ2005} ($P_{\rm constant, joint} < 0.32\%$) to identify a flaring
source. The observed count rate of the flare is about five times the average. As
shown in Figure~\ref{fig:n4697x_med_flr}, the flare is marked by a clearly
increased rate (larger slope in the cumulative light curve) that straddles the
cumulative light curve for a constant rate source. Since the K-S test looks for
the largest deviation from a constant rate, it is not surprising that the K-S
test fails to detect this type of flare. Follow-up of the optical counterpart to
Source 155 is necessary to determine if this source is an flaring LMXB in a GC
or a source unrelated to NGC~4697. As discussed earlier, Source 156 is the
foreground star \object{BD-05 3573} that undergoes a clear flare with 99 counts. 
In Figure~\ref{fig:n4697x_lng_flr}, we overlay the flare count rate determined
by the flare detection algorithm for Source 156, which is about twice the
average rate.

If we consider searching for flares over multiple observations, one can
calculate the joint probability the flares are not real, $P_{\rm constant,
joint}$. Thirteen sources in the Analysis Sample have a $P_{\rm constant,
joint}$ that indicate a $> 2 \sigma$ detection of flaring across multiple
observations; however, the detected flares in two of those sources (137 and 146)
have disparate enough durations that their separate flare detections probably
should not be combined. Only 4 sources (57, 71, 75, and 156) have flaring which
is significant at the $> 3 \sigma$ level across multiple observations; these are
the sources discussed in \citepalias{SSJ2005}, plus the foreground star
\object{BD-05 3573}. After correcting for the number of sources expected from
statistical fluctuations, we find that $4.2^{+3.3}_{-2.5}\%$ of the Analysis
Sample sources exhibit flaring which is detectable at the $> 2 \sigma$ level
with our new algorithm. This is consistent with the $3.1^{+2.4}_{-1.5}\%$ of
the Analysis Sample sources that exhibit flaring which is detectable at the
$> 3 \sigma$ level.

\section{Interobservation Variability}
\label{sec:n4697x_src_var_inter}

Our observations of NGC~4697 are among the first designed to measure
interobservation variability with strong coverage over a wide time baseline. 
Some variability results have been presented before (e.g., Centaurus A:
\citealt{KKF+2001}, M87 \citealt{JCF+2004,Ir2006}, NGC 1399
\citealt{LAM2005,Ir2006}, and NGC 4636: \citealt{PRF+2008}); however, these
results did not have the same quality temporal coverage as our NGC~4697
observations. More recently, variability results were discussed for NGC~3379
\citep{BFK+2008}, with a similar coverage of time baselines as in our NGC~4697
observations. We searched for luminosity variability between any two
observations, as well as luminosity or spectral variability over the entire set
of observations.

\subsection{Luminosity Variability Between Two Observations}
\label{sec:n4697x_src_var_lum2}

\begin{figure}
\plotone{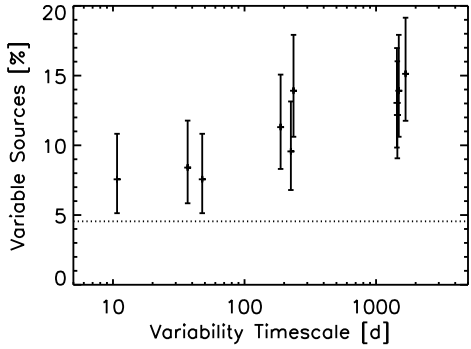}
\caption[Percentages of Variable Discrete X-ray Sources in NGC~4697]{
Percentage of Analysis Sample sources that appear variable (at the
$> 2 \sigma$ level) between two observations against the time period
between the observations.
Note that the error bars are not independent as the pairs contain the
same five observations.
The dotted horizontal line represents the
percentage of sources that would appear variable just due to
statistical fluctuations.
Although there is a rough tendency for more variability on longer
timescales, a constant variability percentage of 11.3\% is statistically
consistent with the data.
\label{fig:n4697x_2pt}}
\end{figure}

We first searched for variability between pairs of observations by calculating
the value of $\chi^2$ assuming a constant flux. Sources were considered possibly
variable if $\chi^2 >4$, i.e., a $2 \sigma$ detection of variability. This
method differs slightly from \citet{PRF+2008}, which compares the individual
luminosities to each other. The probability that the variability is due to a
statistical fluctuation is listed in columns (2)--(11) of
Table~\ref{tab:n4697x_intervar} for the ten pairs of observations. In
Figure~\ref{fig:n4697x_2pt}, we display the percentage of variable sources as a
function of the timescale between two observations for all of the Analysis
Sample sources that are in the FOV for both observations in each pair. We also
display the percentage of variable sources expected due to statistical
fluctuations at the $2 \sigma$ level. There is a slight tendency for a higher
percentage of variable sources at longer timescales; however, this is not
statistically significant, based on a $\chi^2$ test. (There are two caveats with
this: the error bars in Figure~\ref{fig:n4697x_2pt} are not independent, and
there may be a slight excess variability on the four longest timescales due to
the improper correction of QE degradation for SSs. This latter effect comes
about because we have assumed an average spectrum when correcting for QE
degradation; however SSs have a much softer spectrum that is more strongly
affected by the QE degradation and the differences due to QE degradation will be
most evident in comparisons to the first observation. We estimate that this
latter effect is well within the displayed errors that arise from counting
statistics.) If we treat the error bars as independent, we find that
$11.3\pm1.0\%$ of the Analysis Sample sources are variable on average at the $2
\sigma$ level. Given the expected number of falsely identified variable sources,
$6.7\pm1.0\%$ of the Analysis Sample sources are intrinsically variable between
any pair of observations. If we only consider sources variable at the $3 \sigma$
level, the fraction of variable sources is $6.0\pm0.7\%$. Thus, over timescales
of $\sim 10$--$2000$ days, $\sim 6\%$ of sources are significantly variable. This
number is roughly consistent with that found for Centaurus A
($9.8^{+4.5}_{-3.3}\%$) for a variability timescale of $164 {\rm \, d}$
\citep{KKF+2001}; although they derive a much higher percentage when they
include a generous definition of transient candidates ($14.2^{+2.6}_{-2.3}\%$). 
Our derived variability percentage is lower than that found for a $1115 {\rm \,
d}$ timescale in NGC~4636 \citep[$23.8^{+3.3}_{-3.0}\%$;][]{PRF+2008}; however,
it is consistent with the variability for a day timescale
($4.6^{+3.5}_{-2.2}$\%).

\subsection{Long-Term Luminosity Variability}
\label{sec:n4697x_src_var_lum5}

We can apply a similar $\chi^2$ technique to search for sources that show any
luminosity variability over the entire period of five observations. We refer to
these sources as Long-Term Luminosity Variables (LTLVs), and the probability
$P_{L,{\rm All}}$ that the variability arises from statistical fluctuations is
given in column 12 of Table~\ref{tab:n4697x_intervar}. The difference in
$\chi^2$ compared to a single merged luminosity and combined with the
appropriate degrees of freedom (one less than the number of observations a
source is in the FOV) is used to calculate the probability an observed LTLV is
due to a statistical fluctuation. Sources 148, 154, and 157 are excluded from
these calculations as they are only in the FOV of one observation. From the
Analysis Sample, 26/124 sources ($21.0^{+4.3}_{-3.8}\%$) are LTLVs at the $2
\sigma$ significance level, while $11.3^{+3.6}_{-2.9}\%$ of Analysis Sample
sources are LTLVs at the $> 3 \sigma$ significance level. Statistically
correcting for falsely identified LTLVs, we find $16.4^{+4.3}_{-3.8}\%$ of the
significantly detected sources show LTLV behavior at the $> 2 \sigma$
significance level. Three of these sources are not associated with NGC~4697:
Source 45 has an optical ID that is not a GC and thus is probably due to a
background AGN, Source 117 is a known background AGN, and Source 156 is a known
foreground star.
Two sources (25 and 155) are associated with optical counterparts that may or
may not be GCs in NGC~4697. Sources 39, 41, 51, 77, and 79 are GC-LMXBs that
exhibit LTLV behavior. (Sources 98 and 114 are associated with Kavelaars GC
candidates; new HST-ACS data will be used to test whether these sources are
really GC-LMXBs.) After correcting for the likely percentage of falsely
identified LTLVs, $10.1^{+8.7}_{-6.2}\%$ of GC-LMXBs and $<8.5\%$ of Field-LMXBs
exhibit LTLV behavior at the $> 2 \sigma$ significance level.

The recent analysis of sources in NGC~3379 found that $42\pm5\%$ of their LMXBs
exhibited LTLV behavior \citep{BFK+2008} using a $\chi^2$ test similar to the
one used in this paper; however, they only required a source have a reduced
$\chi^2 > 1.2$. For sources in 2--5 observations, this corresponds to a
$\sim30\%$ false detection rate. After correcting for these false detections,
the rate for LTLVs in NGC 4697 ($\sim 11\%$) and NGC 3379 ($\sim 12\%$) are
comparable.

Since some Galactic LMXBs are known to undergo transitions between
distinct luminosity states, we developed a method that groups observations
together into possible luminosity states.
For every LTLV, we first considered all possible combinations of observation
groupings that divide into two separate states.
That is, we assumed that the luminosity had two distinct values, and that
one or the other of the two values applied to each observation of that
source.
The two luminosity values were allowed to vary, and each observation was
assigned to one of the two luminosity states so as to minimize $\chi^2$.
Some sources still had a much larger $\chi^2$ than the number of dofs.
Given the new $\chi^2$, we could calculate the probability that
any deviations from our two-state hypothesis were due to statistical
fluctuations, given the new number of dofs and the number of combinations
of states. If this probability was low ($<4.6$\%), then we
tried to divide the observations into three luminosity states.
No source required more than three states to be fit acceptably.
We list the states in Table~\ref{tab:n4697x_state}.
For each
state, we list the observations included in a state (column 2), and
the combined luminosity and hardness ratios of the state (columns
3--6). Unless otherwise noted, we placed no requirement on the
temporal pattern of the states; LTLVs whose states
are interspersed temporally may be switching between those states or
the intrinsic variability may be poorly characterized by a simple transition
between states.
There are several categories of luminosity state
transitions that indicate potentially interesting transitions and a
few sources that bear individual scrutiny.

Sources 25, 78, and 110 are SSs that exhibit possible LTLV.
As SSs, they are more affected by QE degradation than the typical
source. Among these sources, the first two have a luminosity that
drops precipitously in the last four observations compared to the
first observation. Since the last four observations all have a similar
amount of QE degradation, we reran the LTLV detection
method only considering the last four observations; neither of the
sources remained LTLVs. Therefore, we do not
consider Sources 25 and 78 as LTLVs.
Since Source 110 does not turn on until the last observation, the QE
degradation does not complicate its identification as an LTLV.

Sources 68 and 117 were the only sources broken into three luminosity states.
The individual luminosities of Source 68 are
qualitatively described by an increase over the first three
observations, followed by the source turning off, and perhaps
beginning to turn on again in the last observation. Its two-state mode
is consistent with this qualitative picture, containing a
higher-luminosity state of the first three observations and a state
where the source is not significantly detected in the last two
observations; however, the probability the two-state description is
correct is $6.1\times10^{-3}$. In its best
three-state mode,
Observations 0784 and 4730 are grouped together in a lower-luminosity
state that is still significantly detected; we note that the
three-state set where the first two observations share a
lower-luminosity state, the third observation is the higher-luminosity
state, and the last two observations are a state where the source is
not significantly detected has a $\chi^2$ only 0.46 higher than the
best three-state set. The individual luminosities of Source 117 vary
significantly with no clear temporal pattern.  In its two-state set,
Observation 4729 is in the lower-luminosity state and the rest of the
observations are in the higher-luminosity state; however, the
probability the two-state description is correct is $4.2\times10^{-5}$.

Some sources had a lower-luminosity state that was consistent with no
emission at the $3 \sigma$ level, and a higher-luminosity state, whose
observations were sequential and had emission detected at the $> 3
\sigma$ level.  We labeled sources following this pattern of states
as transient candidates. For Sources 68 and 117, we considered their two-state
properties for discussions of transience. Eleven of the Analysis Sample sources
($8.9^{+3.3}_{-2.6}\%$) are transient candidates (Sources 11, 37, 41, 45, 51,
67, 68, 77, 94, 110, and 121). For the eleven transient candidate sources we
also calculated the minimum ratio of fluxes at the $3 \sigma$ level. Only the
flux ratio of Source 110 is above 10. We believe Source 110 is a truly transient
source and in \S~\ref{sub:n4697x_source110} we discuss it in detail. The other
transient candidates are either transient sources or LTLVs that have a lower
luminosity state that is below the detection limit. All eleven transient
candidates appeared to either turn on or turn off during the 4.6 years over
which we observed; no source turned on and then off, or off and then came back
on again. If we assume all the LMXBs in NGC~4697 are long-term transients, and
that our detected transient sources are just those which turned off or on during
our observation period, then the mean outburst duration for LMXBs in NGC~4697 is
$103^{+42}_{-28} {\rm \, yr}$. Under this hypothesis, the duty cycle of LMXBs
would also depend on their recurrence timescale. Since our observations find no
recurring transients, we have adopted the crustal heating recurrence timescale
for the Galactic NS-LMXB KS~1731$-$260 of $\sim 1500 {\rm \, yr}$
\citep{RBB+2002}. (The adoption of a crustal reheating timescale also implies
that we are assuming all the LMXBs in NGC~4697 are NS-LMXBs. While this is not
necessarily true, we note that the recurrence timescale of Galactic long-term
transient BH-LMXBs is unknown. As such, we present this calculation for
illustrative purposes only.) We note that this implies a duty cycle of $\sim
7\%$, which is comparable to the duty cycle obtained comparing active and
quiescent LMXBs in Galactic GCs \citep[$\sim 12\%$;][]{HGL+2003}. We also note
that the number of transient candidates and possible transient candidates in
NGC~3379 yields similar duty cycles \citep{BFK+2008}, although their study
uses a different definition of transience.

Six of the LTLVs have a $> 2 \sigma$ difference in at least one of
their hardness ratios between their states (Sources 11, 45, 117, 121,
134, and 156; Table~\ref{tab:n4697x_state}, columns 4--6).  These
sources appear to exhibit luminosity/spectral state transitions. Since
the lower-luminosity state in Sources 45 and 121 is not even a
1$\sigma$ significant detection, we exclude these sources from this
discussion. Although Source 11 is a Field-LMXB that is a transient
candidate, it may also be a luminosity/spectral state transition
source. The luminosity of the lower-luminosity state (Observation
4730) is more than $2 \sigma$ significant and appears to be
3.4$\sigma$ softer in $H_{21}$, than the higher-luminosity state.
Although this softer/fainter to harder/brighter transition is opposite
of the conventional relationship in BH state transitions, it is
possible that this discrepancy results from the softer bands used to
calculate the hardness with {\it Chandra} (Juett et al.\ 2008 in
preparation). Alternatively, this source may be transitioning from the
thermal state (formerly called high/soft) to the steep power law state
(formerly called very high). For example, in its 1996--1997 outburst,
the Galactic LMXB \object{GRO J1655-40} is softer and fainter in its
thermal state compared to its steep power law state \citep{RM2006}.
Source 117, a known AGN, appears to display the more typical
harder/fainter to softer/brighter transition \citep[e.g., the spectral
correlation of Seyfert galaxy NGC~4151 in][]{PPA+1986}; Observation
4729 is $\sim 4$--5$\sigma$ harder in $H_{31}$ and $H_{32}$ than the
two higher luminosity states. Its spectrum seems to change most in the
2--6 keV band. Source 134, a suspected AGN based on its hardness
ratios, appears to undergo a change in its 1--2 keV band. The
lower-luminosity state is $2.2 \sigma$ harder in $H_{21}$, but $2.0
\sigma$ softer in $H_{32}$ than its higher-luminosity state.  Finally,
source 156, a known foreground star, undergoes a transition similar to
source 11. The lower-luminosity state of source 156 is $2.4 \sigma$
softer in $H_{21}$ than its higher-luminosity state.
This behavior is not surprising given that stellar flares generally heat their
corona \citep{Gu2004} and the higher-luminosity state occurs in the observation
where this source flares.

\subsection{Long-Term Hardness Ratio Variability}
\label{sec:n4697x_src_var_hr}

Our method for identifying LTLV can also be applied to
hardness ratios;
we define sources whose hardness ratios vary as Long Term Hardness Variables
(LTHVs).
Sources that have variations in the hardnesses ratios that are significant
at the $2 \sigma$ level have the probabilities that their variations are
due to statistical fluctuations listed in Table~\ref{tab:n4697x_intervar},
columns 13--15.
We identify fewer LTHVs than LTLVs
because the errors in hardness ratios are larger than in
luminosities.
Therefore, we only summarize the results for
the 83 sources in multiple observations and with luminosities detected
at the 5$\sigma$ level. We detect $6.0^{+3.9}_{-2.6}\%$,
$7.3^{+4.1}_{-2.9}\%$, and $6.1^{+3.9}_{-2.6}\%$ sources that exhibit
$H_{21}$, $H_{31}$, and $H_{32}$ variability at the $> 2 \sigma$
significance level, respectively.  Corrections for falsely identified
variable sources reduce the percentages of hardness variable sources
to $<5.3\%$, $<6.9\%$, and $<5.5\%$.
That is, the numbers of sources selected purely based on hardness ratio
variations could result from statistical fluctuations.

We also applied the same grouping of observations into states
as in \S~\ref{sec:n4697x_src_var_lum5}, using individual hardness ratios
as opposed to luminosities
(Table~\ref{tab:n4697x_state}, lower portions).
Here we discuss sources that were
identified as LTHVs in at least two hardness ratios,
or identified as LTHVs in at least one hardness
ratio and also as LTLVs.  Sources 10, 66, and 102 are in
the first category, while Sources 11, 117, 134, and 156 are in the the
second. We discuss attempts to resolve the difference in their states,
and the implications of those attempts.

For Source 10, the selections by $H_{31}$ and $H_{32}$ agree, pointing
to a probable decrease in the 2--6 keV count fraction for
Observation 4729.  Although the best-fit states disagree for Source
66, matching the $H_{32}$ selected state to the $H_{31}$ selected
state leads to a $\chi^2$ that is only 0.7 higher than the best
$H_{32}$ selected state.  The fraction of 2--6 keV counts is
higher for Observation 4729.  Similarly, matching the $H_{31}$
selected state to the $H_{32}$ selected state leads to a $\chi^2$ that
is only 0.2 higher than the best $H_{31}$ selected state for Source
102; the 2--6 keV count fraction is higher for Observations 4727
and 4729.

Sources 11, 117, 134, and 156 were all identified as
luminosity/spectral state transitions using the LTLV
selection. The hardness ratio two-state sets in Source 117 are
consistent with the two-state luminosity set; the 2--6 keV count
fraction is higher in the lower-luminosity state (Observation 4729). The
$H_{21}$ and luminosity states match for Source 156, where the
lower-luminosity state is softer in $H_{21}$ than the
higher-luminosity state.
Sources 11 and 134 are more difficult to resolve as clear
luminosity/spectral state transitions as the the grouping of observations
into states by luminosity and hardness ratios do not match. If we match the
$H_{21}$ selected state to the luminosity selected state for Source 11, the
$\chi^2$ is 4.1 higher;
however, both sets of states are consistent with the lower-luminosity state
being softer in $H_{21}$. The resolution for Source 134 is even more
problematic. This source appears best described by a non-monotonic evolution of
hardness with luminosity. As the source luminosity increases, it first becomes
softer and then becomes harder in all three hardness ratio bands; however, the
softest observation changes between 4728 or 4729 depending on the hardness
ratio.

\subsection{Source 110: A Transient Supersoft Source}
\label{sub:n4697x_source110}

We consider Source 110 to be a clear transient source. In the sum of the first
four observations, there were 5 counts in the source aperture and 19 counts in
the (three times) larger background aperture. After correcting for QE
degradation of a typical source in NGC~4697, the corresponding rate is $(-1.3
\pm 3.8) \times 10^{-5} {\rm \, cnt \, s}^{-1}$; the source is clearly
undetected. We also calculated the corrected rate including observations
4727-4729, which have a similar level of QE degradation, to be $(-1.0 \pm 4.6)
\times 10^{-5} {\rm \, cnt \, s}^{-1}$. In the final observation, there were 87
counts in the source aperture and 4 counts in the background aperture,
corresponding to a QE degradation corrected rate of $(32.6 \pm 4.0) \times
10^{-4} {\rm \, cnt \, s}^{-1}$. The ratio of the count rate in Observation 4730
to the other observations is $>27$ at the $3 \sigma$ level. Therefore, we only
consider data from Observation 4730 hereafter. Within the 0.3--$6.0 {\rm \,
keV}$ band, only two of the source aperture photons were above $1{\rm \, keV}$
(1.03 and $3.9 {\rm \, keV}$). Source 110 is clearly supersoft. Its $3\sigma$
limits on observed hardness ratios are $H_{21} < -0.89$ and $H_{31} < -0.88$. 
Analysis of our flanking-field HST-ACS observations indicates Source 110 is not
associated with an optical source, although it is about $1\farcs9$ away from a
GC candidate. From simulated point sources at the location of Source 110 we
placed 90\% confidence limits on magnitudes for a single HST-ACS image of $g >
26.3$ and $z > 25.3$ in AB mag. Since Source 110 is in an overlapping region of
two fields, these magnitudes correspond to the 99\% confidence limit on an
optical counterpart to Source 110.

Since Source 110 is both transient and SS, we explored its
spectrum in greater detail.  We examined both the raw spectrum and a
binned spectrum. Since Source 110 has only 87 gross counts in Observation
4730, we created five bins with at least 16 counts in the 0.3--$1 {\rm \, keV}$
range and a bin with two counts from 1--$6 {\rm\, keV}$.
These bins were weighted by Gehrels errors.

Luminous SSs in our Galaxy and M31 are generally believed to be
accreting WDs. Such sources are often fit by a blackbody (bb) model.
Compared to the best-fit for this model, the spectrum
appears to rise too quickly within the 0.3--$1 {\rm \, keV}$ band to
be consistent with Galactic absorption in the raw spectrum. In
the binned spectrum, the best-fit model with Galactic absorption was
strongly rejected, $\chi^2 = 16.78$ for 4 dof ($kT_{\rm bb} = 129\pm16
{\rm \, eV}$). This model is also hotter than typical fits to SS-WDs
\citep[15--$80 {\rm \, eV}$;][]{KH1997}. If we allow the absorption to
vary, we do get a high absorbing column and lower temperature,
$N_H = 1.2^{+1.0}_{-0.6} \times 10^{22} {\rm \, cm}^{2}$ and
$kT_{\rm bb} = 55^{+26}_{-19} {\rm \, eV}$
(90\% individual confidence limits), for
an acceptable fit with $\chi^2 = 2.04$ for 3 dof.
However, the implied X-ray and bolometric luminosity in the
0.3--$10 {\rm \, keV}$,
$L_X = 3.0 \times 10^{41} {\rm \, ergs \, s}^{-1}$
and
$L_{\rm bol} = 1.6 \times 10^{42} {\rm \, ergs \, s}^{-1}$
are much too large for a WD at the distance of NGC~4697.
Even the lowest X-ray and bolometric luminosities allowed by the uncertainties
at 90\% confidence for jointly varying absorption and temperature, $L_X = 1.6
\times 10^{39} {\rm \, ergs \, s}^{-1}$ and $L_{\rm bol} = 3.0 \times 10^{39}
{\rm \, ergs \, s}^{-1}$ , are much too luminous for a SS-WD in NGC~4697. These
luminosities come at hotter temperatures, $90 {\rm \, eV}$, than seen in SS-WDs.
Although more detailed WD atmosphere models can lead to luminosities lower by an
order of magnitude than derived from the bb model
\citep[e.g.,][]{HTK1994,BKO1998,OZM+2007}, they also tend to increase
the temperature. Thus, we consider it unlikely that Source 110 is a SS-WD in NGC
4697. On the other hand, our optical limit suggests that it is unlikely that
Source 110 is a SS-WD within the Milky Way. Of the likely companions for a
SS-WD, main sequence stars earlier than F5, red giants overflowing their Roche
lobe, and asymptotic giant branch stars with winds \citep{KH1997}, an F5
main-sequence star is the faintest with $M_{V} \sim 3.5$ \citep[pg.\
107][]{BM1987}. We would detect such a source out to $\sim 300 {\rm \, kpc}$ in
our HST images. Thus, Source 110 is unlikely to be a SS-WD, either as a
foreground Galactic source or a source in NGC~4697.

A very soft, luminous X-ray transient with an excess absorption column
has been observed in M101 \citep[M101~ULX-1;][]{KD2005}. The
combination of spectra, outburst luminosity, and transience was used
to suggest that source is an intermediate-mass BH (IMBH).  The disk
blackbody model \citep[hereafter diskbb,][]{MIK+1984} is a multi-color
blackbody model that is often used to represent BH soft X-ray
transients (SXTs), which have $kT_{\rm diskbb} < 1.2 {\rm \, keV}$
\citep{TS1996}.
As with the simple blackbody model, the spectrum is poorly fit by a
model with Galactic absorption; $kT_{\rm diskbb} = 163\pm27 {\rm \,
eV}$ yields $\chi^2 = 20.05$ for 4 dof. If we allow the absorption to
vary, the absorption and temperature are strongly correlated and span
a fairly large range, but the fit is much improved with a
$\chi^2 = 2.02$ for 3 dof.  We find $N_H = 1.2^{+1.0}_{-0.6} \times 10^{22}
{\rm \, cm}^{2}$ and $kT_{\rm diskbb} = 59^{+28}_{-22} {\rm \, eV}$
(90\% individual confidence limits). The 0.3--$10 {\rm \, keV}$ X-ray
luminosity of acceptable spectral models (within the 90\%
two-dimensional confidence interval for varying absorption and
temperature) ranges from $3.5 \times 10^{39}$ -- $6.8 \times 10^{46}
{\rm \, ergs \, s}^{-1}$.
The bolometric luminosity range is even
larger, $1.4 \times 10^{40} < L_{\rm bol} < 4.8 \times 10^{48}
{\rm \, ergs \, s}^{-1}$.
We use the normalization of the diskbb
model to estimate a BH mass following equation (3) in \citet{FRM2004}.
Following \citet{FRM2004}, we also assume a correction factor of $f_0 = 1.35$,
an inclination angle of $i = 60\degr$,
and an innermost disk radius of $6 \, G \, M_{\rm BH}/c^2$, where
$M_{\rm BH}$ is the BH mass.
Dividing our derived bolometric luminosity by the Eddington luminosity,
$L_{\rm Edd} = 1.26 \times 10^{38} ( M_{\rm BH} / M_{\odot} ) {\rm \, ergs \,
s}^{-1}$,
gives the Eddington efficiency of the accretion disk, which we use 
later in evaluating the plausibility of the model.
The value of the disk blackbody temperature also provides a second estimate
of the black hole mass, which is roughly
$M_{\rm BH} / M_{\odot} \sim ( kT_{\rm diskbb} / 1 \, {\rm keV} )^{-4}$
\citep{MKM+2000}.
Considering
models within the 90\% two-dimensional confidence interval for varying
absorption and temperature, these two estimates of BH mass agree to
within a factor of two for absorption columns between 0.5 and $1\times
10^{22} {\rm \, cm}^{2}$.
Within this region, our source is consistent
with BH masses of
$\sim ( 7.2 \times 10^{3}$--$1.1\times 10^{6} ) \, M_\odot$
accreting at $\sim 1.6$--4.7\% of the Eddington limit.
Thus, this source is consistent with a rather massive IMBH or a small
supermassive black hole about a factor of 2--3 smaller in mass than the
one at the center of the Milky Way \citep{GMB+2000}.
Understanding the origin of such a large BH outside of the center of
a galaxy or of a GC is a formidable challenge to our current theories of BH
formation.

Some AGNs have ultrasoft spectra. For example, \citet{PMC+1992}
identified 53 ultrasoft AGN candidates with {\it Einstein
Observatory}.  These sources had 0.16--$0.56 {\rm \, keV}$ count rates
nearly three times that of their 0.56--$1.08 {\rm \, keV}$ count
rates.  This corresponds to steep power-law photon indices, $\Gamma
\gtrsim 3$, with Galactic absorption. In a study of variability in
soft X-ray selected AGNs, \citet{GTB2001} found that AGNs with steeper X-ray
spectra had larger short-term variability than those with flatter spectra. Two
of the transient sources, \object{IC 3599} and \object{HE 0036-5133} (also
called WPVS~007) were also previously reported as ultrasoft. These sources may
be analogs to Source 110. A variety of complicated models have been used to
explain the spectra of AGNs with excess soft emission compared to the power-law
models derived from harder bands \citep[e.g.,][]{GD2004}. Given the relative
paucity of our data on Source 110, we chose to only consider simple power-law
and blackbody models. The power-law model with Galactic absorption, $\Gamma =
4.23^{+0.86}_{-0.58}$, is an exceptionally poor fit, $\chi^2 = 30.20$ for 4 dof. 
Even with a free absorption, power-law models with reasonable photon indices,
$\Gamma < 10$, did not fit our data well, $\chi^2 > 12.25$ for 3 dof. From
above, we know that the disk blackbody model is a good statistical
representation of the spectra. However, our derivation of the BH masses in the
above paragraph was predicated on the source being at the distance of NGC~4697. 
As we allow the source to lie farther behind NGC~4697, the agreement between the
mass derived from the normalization and the mass derived from the temperature
begins to break down. Thus, it is unlikely that the disk blackbody model fully
represents the physical mechanism behind the emission. Regardless of the model
we choose to represent the emission, we measure an absorbed X-ray flux
(0.5--$8.0 {\rm \, keV}$) of $F_X
\sim 7
\times 10^{-15} {\rm \, ergs \, s}^{-1} {\rm \, cm}^{-2}$.
At that flux level, we only expect $\sim 3.6$ sources
unrelated to NGC~4697 for galactocentric radii interior to that of
Source 110.
Thus, the probability Source 110 is a
background AGN is $\sim 3.3\%$ from the X-ray data alone.
If we assume that the transient behavior observed was not an extreme flare that
began between 2004 February 02 and August 18, but ended before the HST-ACS
flanking field observations began in 2005 December 19, we can place an additional
constraint using the logarithm of its X-ray--to--optical flux ratio, $\log (F_X/F_{\rm opt})$,
where $F_{\rm opt}$ is the optical flux in the $z$ band.
Based on our non-detection of a counterpart to Source 110, its
$\log (F_X/F_{\rm opt})$ is greater than 0.9 at the 99\% confidence level.
Such a large ratio is expected only from the most
obscured AGNs. There are two problems with suggesting Source 110 is
an obscured AGN.
First, the most obscured AGNs can have columns one
to two orders of magnitude larger than the estimated column for Source
110.
In fact, its column is actually near the division point typically
used for obscured and unobscured AGN
\citep[$10^{22} {\rm \, cm}^{2}$;][]{TUC+2004}.
Second, very few sources are predicted to have such high X-ray--to--optical flux
ratios. In \citet{TUC+2004}, a model based on the AGN unification paradigm
predicts only $\sim13\%$ of AGNs will have $\log (F_X/F_{\rm opt}) > 0.8$. In
summary, although some of its X-ray properties may be consistent with transient
ultrasoft AGN behavior, the combination of the expected number of background
AGNs with the high X-ray--to--optical flux ratios makes it unlikely (probability
$\sim 0.4\%$) that Source 110 is a transient ultrasoft AGN.

\section{Conclusions}
\label{sec:n4697x_conclusion}
Our multi-epoch {\it Chandra} observations were planned to shed light on LMXB
variability in NGC~4697, the nearest, optically luminous, elliptical galaxy. We
measure variability on timescales of seconds to years. Although we do not detect
any periodic LMXBs at significant levels, we detect five sources with
significant variability on timescales of seconds to hours
(Source 57, 71, 75, 155, and 156).

When we compare the luminosities of sources between every pair of observations,
we determine that $6.7\pm1.0\%$ of the Analysis Sample sources are variable over
timescales of $\sim 10$--$2000$ days. There is a slight increase in the fraction
of variable sources with the variability timescale, but it is not found to be
statistically significant. Our number is roughly consistent with that found for
Centaurus A for a variability timescale of $164 {\rm \, d}$
\citep{KKF+2001} and for NGC~4636 for a day timescale, but is lower
than that found for a $1115 {\rm \, d}$ timescale in NGC~4636 \citep{PRF+2008}.
New multi-epoch observations of NGC~1023, NGC~3379, NGC~4278, NGC~4365, and
Centaurus A will help fill in more variability timescales. As this phase space
is populated, we will be able to test if more sources are variable on longer
timescales than shorter timescales. Increased knowledge of the temporal
evolution of variability will impact models that explore the nature of the
observed LMXBs \citep[e.g.,][]{PB2002}.

We are more sensitive to variability when we consider all five observations
simultaneously. We find that $\approx16\pm4\%$ of Analysis Sample sources do not
have a constant luminosity over all five observations. If we apply a similar
requirement for a source to be considered variable to the \citet{BFK+2008}
analysis of LMXBs in NGC~3379 as in this study, we find that the number of
variable sources is $12\pm5\%$, consistent with what we find for NGC~4697. The
fraction of GC-LMXBs that are variable is statistically
indistinguishable from the fraction of variable Field-LMXBs in our sample due
to the small number of variable sources. To better test for any difference
between the two populations, multi-epoch variability from multiple galaxies will
need to be combined.
 
For the detected sources we group observations into sets of luminosity and
hardness ratio states. Among these variable sources, we identify eleven
transient candidates. If we assume all the LMXBs in NGC~4697 are long-term
transients, we estimate that they are on for $\sim 100 {\rm \, yr}$ and have a
7\% duty cycle. The latter is comparable to the duty cycle obtained comparing
active and quiescent LMXBs in Galactic GCs \citep{HGL+2003}.

The mean outburst duration for LMXBs in NGC~4697 of $\sim 100 {\rm \, yr}$ is
consistent with a lower limit of $\sim 50 {\rm \, yr}$ found for 33 luminous
LMXBs in NGC~1399 and M87 that had X-ray luminosities exceeding $8 \times
10^{38}$ ergs s$^{-1}$ \citep{Ir2006}. These 33 sources were assumed to be
BH-LMXBs because of their sustained super-Eddington luminosities over several
year timescales, and most likely represent extragalactic analogs of the Milky
Way X-ray binary GRS1915+105. Since a majority of the sources in NGC~4697 are
probably NS-LMXBs, there does not appear to be a measurable difference between
the outburst durations of long-term transient NS-LMXBs and BH-LMXBs.
 
We also find four sources with a combined luminosity/spectral transition, only
one of which is likely associated with NGC 4697. One limit to our ability to
detect such sources is the duration of our single observations. Although
observations of NGC~3379 and NGC~4278 have longer single observation durations,
this issue is likely to be best addressed by a recently completed set of six,
100 ks observations of Centaurus A, which is at least 2.5 times closer than any
other massive early-type galaxy.

Source 110 is clearly a supersoft transient source with no detected
counterpart in {\it HST} observations. Although it only has $\approx
85$ X-ray counts, our spectral analysis rules out the possibility that
this source is a luminous supersoft WD. The high X-ray--to--optical flux
ratio and relative small number of expected background AGNs argue
against the source being a transient ultrasoft AGN. Our spectral fit
could be consistent with an accreting BH with a mass in the range of
$\sim$$( 7 \times 10^{3}$--$10^{6} ) \, M_\odot$;
however, the origin of such a large BH
outside of the galaxy center or a GC is a formidable challenge
to our current theories of BH formation. Further observations of
NGC~4697 could increase the information we can extract from the
spectra of Source 110; however, more work must also be done to
increase our understanding of the X-ray spectra of non-stellar-mass
black holes.

\acknowledgements

We thank Kris Beckwith, Marina Orio, and and Roseanne
di Stefano for very helpful discussions.  Support for this work was
provided by NASA through {\it Chandra} Award Numbers GO4-5093X,
AR4-5008X, and GO5-6086X, and through HST Award Numbers
HST-GO-10003.01-A, HST-GO-10582.02-A, HST-GO-10597.03-A, and 
HST-GO-10835.03-A.
G.~R.~S.\ acknowledges the receipt of an ARCS fellowship and support
provided by the
F. H. Levinson Fund.

\bibliography{ms}


\clearpage
\tabletypesize{\footnotesize}
\begin{landscape}
\LongTables

\clearpage
\begin{deluxetable}{lrrrrrrrrrrrrrr}
\setlength{\tabcolsep}{3.78pt}
\tablewidth{0pt}
\tablecaption{Sources Which Vary Between Pairs of Observations
\label{tab:n4697x_intervar}}
\tablehead{
\colhead{Source}&
\colhead{$P_{L,AB}$}&
\colhead{$P_{L,AC}$}&
\colhead{$P_{L,AD}$}&
\colhead{$P_{L,AE}$}&
\colhead{$P_{L,BC}$}&
\colhead{$P_{L,BD}$}&
\colhead{$P_{L,BE}$}&
\colhead{$P_{L,CD}$}&
\colhead{$P_{L,CE}$}&
\colhead{$P_{L,DE}$}&
\colhead{$P_{L,All}$}&
\colhead{$P_{H_{21},All}$}&
\colhead{$P_{H_{31},All}$}&
\colhead{$P_{H_{32},All}$}\\
\colhead{(1)}&
\colhead{(2)}&
\colhead{(3)}&
\colhead{(4)}&
\colhead{(5)}&
\colhead{(6)}&
\colhead{(7)}&
\colhead{(8)}&
\colhead{(9)}&
\colhead{(10)}&
\colhead{(11)}&
\colhead{(12)}&
\colhead{(13)}&
\colhead{(14)}&
\colhead{(15)}
}
\startdata
\phn\phn6               & \cennodata & \cennodata & \cennodata & \cennodata & \cennodata & \cennodata &    1.7E-02 & \cennodata &    1.4E-02 & \cennodata & \cennodata & \cennodata & \cennodata & \cennodata \\
\phn\phn7               & \cennodata &    3.0E-02 & \cennodata & \cennodata &    2.1E-02 & \cennodata & \cennodata &    4.1E-02 & \cennodata & \cennodata & \cennodata & \cennodata & \cennodata & \cennodata \\
\phn\phn8               & \cennodata & \cennodata & \cennodata & \cennodata & \cennodata & \cennodata & \cennodata & \cennodata & \cennodata & \cennodata & \cennodata & \cennodata & \cennodata & \cennodata \\
\phn\phn9               &    3.4E-02 & \cennodata & \cennodata & \cennodata & \cennodata & \cennodata &    1.6E-02 & \cennodata &    3.5E-02 & \cennodata & \cennodata & \cennodata & \cennodata & \cennodata \\
\phn10                  & \cennodata & \cennodata & \cennodata & \cennodata & \cennodata & \cennodata & \cennodata & \cennodata & \cennodata & \cennodata & \cennodata & \cennodata &    5.1E-04 &    2.4E-02 \\
\phn11\tablenotemark{a} & \cennodata & \cennodata & \cennodata &    3.0E-03 & \cennodata & \cennodata &    7.9E-04 & \cennodata &    2.1E-03 &    8.5E-04 &    2.4E-04 &    3.0E-02 & \cennodata & \cennodata \\
\phn15                  & \cennodata & \cennodata &    1.8E-02 & \cennodata & \cennodata & \cennodata & \cennodata & \cennodata & \cennodata & \cennodata & \cennodata & \cennodata & \cennodata & \cennodata \\
\phn19\tablenotemark{b} & \cennodata &    8.9E-03 & \cennodata & \cennodata & \cennodata & \cennodata & \cennodata & \cennodata & \cennodata & \cennodata & \cennodata & \cennodata & \cennodata & \cennodata \\
\phn23                  & \cennodata & \cennodata & \cennodata &    1.5E-03 & \cennodata & \cennodata &    2.7E-04 & \cennodata &    4.7E-04 &    9.4E-03 &    2.5E-02 & \cennodata & \cennodata & \cennodata \\
\phn25\tablenotemark{b} &    3.1E-07 &    7.9E-08 &    1.9E-03 &    5.1E-03 & \cennodata & \cennodata & \cennodata & \cennodata & \cennodata & \cennodata &    7.8E-06 & \cennodata & \cennodata & \cennodata \\
\phn26                  & \cennodata & \cennodata & \cennodata & \cennodata &    6.3E-03 & \cennodata &    2.5E-02 & \cennodata & \cennodata & \cennodata & \cennodata & \cennodata & \cennodata & \cennodata \\
\phn28                  & \cennodata &    2.1E-02 & \cennodata & \cennodata & \cennodata & \cennodata & \cennodata & \cennodata & \cennodata & \cennodata & \cennodata & \cennodata & \cennodata & \cennodata \\
\phn31                  & \cennodata & \cennodata &    2.9E-02 & \cennodata & \cennodata & \cennodata & \cennodata & \cennodata & \cennodata &    3.7E-02 & \cennodata & \cennodata & \cennodata & \cennodata \\
\phn34                  & \cennodata & \cennodata & \cennodata & \cennodata & \cennodata & \cennodata & \cennodata &    4.0E-02 & \cennodata & \cennodata & \cennodata & \cennodata & \cennodata & \cennodata \\
\phn37\tablenotemark{a} & \cennodata & \cennodata & \cennodata &    6.8E-11 & \cennodata & \cennodata &    6.1E-07 & \cennodata &    1.9E-06 &    7.6E-04 &    2.8E-05 & \cennodata & \cennodata & \cennodata \\
\phn39                  &    1.3E-05 &    3.7E-06 &    2.5E-04 &    1.7E-04 & \cennodata & \cennodata & \cennodata & \cennodata & \cennodata & \cennodata &    6.8E-05 & \cennodata & \cennodata & \cennodata \\
\phn41\tablenotemark{a} &    2.6E-03 &    2.8E-02 & \cennodata & \cennodata & \cennodata & \cennodata & \cennodata & \cennodata & \cennodata & \cennodata &    2.7E-02 & \cennodata & \cennodata & \cennodata \\
\phn45                  &    1.4E-05 &    3.4E-05 &    1.1E-04 &    3.3E-08 & \cennodata & \cennodata & \cennodata & \cennodata & \cennodata & \cennodata &    3.0E-04 & \cennodata & \cennodata & \cennodata \\
\phn48                  & \cennodata & \cennodata & \cennodata & \cennodata & \cennodata & \cennodata & \cennodata & \cennodata & \cennodata & \cennodata & \cennodata & \cennodata &    6.5E-04 & \cennodata \\
\phn51\tablenotemark{a} & \cennodata & \cennodata & \cennodata &    1.1E-02 & \cennodata &    2.0E-02 &    2.5E-04 &    1.0E-03 &    4.7E-07 & \cennodata &    9.2E-06 & \cennodata & \cennodata & \cennodata \\
\phn52                  & \cennodata & \cennodata & \cennodata & \cennodata & \cennodata &    2.3E-02 & \cennodata &    2.0E-02 & \cennodata & \cennodata & \cennodata & \cennodata & \cennodata & \cennodata \\
\phn53                  & \cennodata & \cennodata & \cennodata & \cennodata &    8.7E-03 & \cennodata & \cennodata & \cennodata & \cennodata & \cennodata & \cennodata & \cennodata & \cennodata & \cennodata \\
\phn58                  & \cennodata & \cennodata & \cennodata & \cennodata & \cennodata & \cennodata & \cennodata & \cennodata & \cennodata & \cennodata & \cennodata & \cennodata &    1.5E-02 & \cennodata \\
\phn59                  & \cennodata & \cennodata & \cennodata & \cennodata & \cennodata & \cennodata & \cennodata & \cennodata & \cennodata &    2.9E-02 & \cennodata & \cennodata & \cennodata & \cennodata \\
\phn60                  & \cennodata & \cennodata & \cennodata &    1.2E-02 & \cennodata & \cennodata & \cennodata & \cennodata & \cennodata &    2.3E-02 & \cennodata & \cennodata & \cennodata & \cennodata \\
\phn63                  &    8.8E-03 & \cennodata & \cennodata & \cennodata & \cennodata & \cennodata & \cennodata & \cennodata & \cennodata & \cennodata & \cennodata &    2.9E-02 & \cennodata & \cennodata \\
\phn66                  & \cennodata & \cennodata & \cennodata & \cennodata & \cennodata & \cennodata & \cennodata & \cennodata & \cennodata & \cennodata & \cennodata & \cennodata &    1.5E-02 &    2.5E-03 \\
\phn67\tablenotemark{a} &    3.9E-06 & $<$1.0E-15 &    6.4E-09 &    8.4E-06 & \cennodata & \cennodata & \cennodata & \cennodata & \cennodata & \cennodata & $<$1.0E-15 & \cennodata & \cennodata & \cennodata \\
\phn68\tablenotemark{a} & \cennodata &    1.4E-03 &    3.1E-12 &    4.3E-02 &    3.2E-02 & $<$1.0E-15 &    2.9E-03 & $<$1.0E-15 &    5.4E-08 &    4.1E-03 & $<$1.0E-15 & \cennodata & \cennodata & \cennodata \\
\phn69                  & \cennodata & \cennodata & \cennodata & \cennodata &    3.1E-02 & \cennodata & \cennodata & \cennodata & \cennodata & \cennodata & \cennodata & \cennodata & \cennodata & \cennodata \\
\phn75                  & \cennodata & \cennodata & \cennodata & \cennodata & \cennodata & \cennodata &    1.0E-02 & \cennodata &    1.2E-02 &    2.9E-03 & \cennodata & \cennodata & \cennodata & \cennodata \\
\phn77\tablenotemark{a} & \cennodata &    7.0E-10 &    2.4E-04 &    3.6E-13 &    6.8E-05 &    2.5E-02 &    3.3E-06 & \cennodata & \cennodata & \cennodata &    4.3E-15 & \cennodata & \cennodata & \cennodata \\
\phn78\tablenotemark{b} &    1.3E-05 &    5.9E-07 &    2.1E-05 &    8.5E-05 & \cennodata & \cennodata & \cennodata & \cennodata & \cennodata & \cennodata &    1.1E-04 & \cennodata & \cennodata & \cennodata \\
\phn79                  & \cennodata & \cennodata & \cennodata & \cennodata & \cennodata & \cennodata &    2.8E-04 & \cennodata & \cennodata & \cennodata &    3.4E-02 & \cennodata & \cennodata & \cennodata \\
\phn81                  & \cennodata & \cennodata & \cennodata & \cennodata & \cennodata & \cennodata & \cennodata & \cennodata & \cennodata & \cennodata & \cennodata & \cennodata & \cennodata &    3.5E-02 \\
\phn94\tablenotemark{a} & \cennodata & \cennodata & \cennodata &    1.6E-03 & \cennodata & \cennodata &    1.9E-04 &    1.6E-02 &    7.6E-06 & \cennodata &    6.9E-04 & \cennodata & \cennodata & \cennodata \\
\phn97                  & \cennodata &    2.5E-02 & \cennodata & \cennodata & \cennodata & \cennodata & \cennodata & \cennodata & \cennodata & \cennodata & \cennodata & \cennodata & \cennodata & \cennodata \\
\phn98                  &    7.1E-04 & \cennodata &    3.6E-02 & \cennodata & \cennodata & \cennodata &    2.5E-03 & \cennodata & \cennodata & \cennodata &    3.8E-03 & \cennodata & \cennodata & \cennodata \\
\phn99                  & \cennodata & \cennodata & \cennodata & \cennodata &    1.4E-02 & \cennodata & \cennodata & \cennodata & \cennodata & \cennodata & \cennodata & \cennodata & \cennodata & \cennodata \\
 100                  & \cennodata & \cennodata & \cennodata & \cennodata & \cennodata & \cennodata & \cennodata & \cennodata & \cennodata & \cennodata & \cennodata & \cennodata & \cennodata & \cennodata \\
 101                  & \cennodata & \cennodata &    4.3E-02 & \cennodata & \cennodata &    4.5E-03 & \cennodata &    1.9E-02 & \cennodata &    5.8E-03 & \cennodata & \cennodata & \cennodata & \cennodata \\
 102                  & \cennodata & \cennodata & \cennodata & \cennodata & \cennodata & \cennodata & \cennodata & \cennodata & \cennodata & \cennodata & \cennodata & \cennodata &    4.1E-02 &    2.4E-02 \\
 103                  & \cennodata & \cennodata & \cennodata & \cennodata & \cennodata & \cennodata & \cennodata & \cennodata &    2.9E-02 &    1.8E-02 & \cennodata &    2.7E-04 & \cennodata & \cennodata \\
 104                  & \cennodata & \cennodata & \cennodata &    8.1E-04 & \cennodata & \cennodata &    2.4E-03 & \cennodata &    2.7E-03 &    6.2E-04 &    2.9E-02 & \cennodata & \cennodata & \cennodata \\
 106                  & \cennodata & \cennodata &    3.6E-02 & \cennodata & \cennodata & \cennodata & \cennodata & \cennodata & \cennodata & \cennodata & \cennodata & \cennodata & \cennodata & \cennodata \\
 108                  & \cennodata & \cennodata & \cennodata &    1.4E-02 & \cennodata & \cennodata & \cennodata & \cennodata & \cennodata & \cennodata & \cennodata & \cennodata & \cennodata & \cennodata \\
 110\tablenotemark{a} & \cennodata & \cennodata & \cennodata & $<$1.0E-15 & \cennodata & \cennodata & $<$1.0E-15 & \cennodata & $<$1.0E-15 & $<$1.0E-15 & $<$1.0E-15 & \cennodata & \cennodata & \cennodata \\
 114                  & \cennodata & \cennodata & \cennodata &    9.7E-03 & \cennodata & \cennodata &    3.4E-02 & \cennodata & \cennodata &    5.6E-03 &    2.0E-02 & \cennodata & \cennodata & \cennodata \\
 115                  &    9.9E-03 & \cennodata & \cennodata & \cennodata & \cennodata & \cennodata & \cennodata & \cennodata & \cennodata & \cennodata & \cennodata & \cennodata & \cennodata & \cennodata \\
 117                  &    1.0E-03 & \cennodata &    8.0E-15 &    2.1E-02 &    1.6E-06 & $<$1.0E-15 & \cennodata &    3.5E-08 &    1.1E-04 & $<$1.0E-15 & $<$1.0E-15 & \cennodata &    5.3E-04 &    1.8E-05 \\
 121\tablenotemark{a} &    1.1E-04 &    4.1E-03 &    2.4E-04 &    8.2E-05 & \cennodata & \cennodata & \cennodata & \cennodata & \cennodata & \cennodata &    2.9E-03 & \cennodata & \cennodata & \cennodata \\
 122                  &    9.8E-04 & \cennodata & \cennodata & \cennodata & \cennodata & \cennodata & \cennodata & \cennodata & \cennodata & \cennodata &    3.6E-02 & \cennodata & \cennodata & \cennodata \\
 124                  & \cennodata & \cennodata & \cennodata &    3.9E-02 & \cennodata & \cennodata & \cennodata & \cennodata & \cennodata & \cennodata & \cennodata & \cennodata & \cennodata & \cennodata \\
 125                  & \cennodata & \cennodata & \cennodata & \cennodata & \cennodata & \cennodata & \cennodata & \cennodata & \cennodata &    6.9E-03 & \cennodata & \cennodata & \cennodata & \cennodata \\
 130                  & \cennodata &    1.4E-03 & \cennodata &    5.1E-04 & \cennodata & \cennodata & \cennodata & \cennodata & \cennodata & \cennodata &    3.7E-03 & \cennodata & \cennodata & \cennodata \\
 131                  & \cennodata & \cennodata & \cennodata &    4.4E-03 & \cennodata & \cennodata &    2.1E-04 & \cennodata & \cennodata & \cennodata &    1.3E-02 & \cennodata & \cennodata & \cennodata \\
 132                  & \cennodata &    1.5E-02 & \cennodata & \cennodata & \cennodata & \cennodata & \cennodata & \cennodata & \cennodata & \cennodata & \cennodata & \cennodata & \cennodata & \cennodata \\
 133                  &    4.1E-02 & \cennodata & \cennodata & \cennodata & \cennodata & \cennodata & \cennodata & \cennodata & \cennodata & \cennodata & \cennodata & \cennodata & \cennodata & \cennodata \\
 134                  &    6.0E-04 & \cennodata &    3.0E-02 & \cennodata & \cennodata & \cennodata &    1.1E-02 & \cennodata & \cennodata & \cennodata &    8.9E-03 &    9.9E-03 &    1.0E-03 &    2.0E-02 \\
 140                  & \cennodata & \cennodata & \cennodata & \cennodata & \cennodata & \cennodata & \cennodata & \cennodata & \cennodata & \cennodata & \cennodata &    3.5E-02 & \cennodata & \cennodata \\
 142                  & \cennodata & \cennodata &    3.8E-02 & \cennodata & \cennodata & \cennodata & \cennodata & \cennodata & \cennodata & \cennodata & \cennodata & \cennodata & \cennodata & \cennodata \\
 144                  & \cennodata & \cennodata & \cennodata & \cennodata &    4.0E-02 & \cennodata &    9.6E-03 & \cennodata & \cennodata &    1.9E-02 &    4.2E-02 & \cennodata & \cennodata & \cennodata \\
 147                  & \cennodata & \cennodata & \cennodata & \cennodata & \cennodata &    1.3E-04 & \cennodata &    7.8E-03 & \cennodata & \cennodata &    4.8E-03 & \cennodata & \cennodata & \cennodata \\
 151                  & \cennodata & \cennodata & \cennodata & \cennodata & \cennodata & \cennodata & \cennodata & \cennodata & \cennodata & \cennodata & \cennodata & \cennodata & \cennodata &    1.4E-02 \\
 152                  & \cennodata & \cennodata & \cennodata & \cennodata & \cennodata & \cennodata & \cennodata & \cennodata & \cennodata & \cennodata & \cennodata &    3.9E-03 & \cennodata & \cennodata \\
 155                  & \cennodata & \cennodata & \cennodata & \cennodata & \cennodata &    2.2E-03 & \cennodata &    1.9E-02 & \cennodata & \cennodata &    7.7E-03 & \cennodata & \cennodata & \cennodata \\
 156                  & \cennodata & \cennodata & \cennodata & \cennodata &    1.1E-07 &    9.0E-04 & \cennodata & \cennodata & \cennodata & \cennodata &    8.5E-07 &    4.2E-02 & \cennodata & \cennodata \\
\enddata
\tablecomments{Observations 0784, 4727, 4728,
4729, 4730 are indicated by A, B, C, D, and E, respectively.
The subscripts to the probability, $P$, indicate the parameter (luminosity $L$
or hardness ratio) and the observations a source is variable over,
while the value of $P$
indicates the probability that the observed variability is due to a
statistical fluctuation.}
\tablenotetext{a}{Source is a transient candidate.}
\tablenotetext{b}{Variations in the QE degradation may be misinterpreted as
variability when comparing Observation 0784 to Cycle 5 observations for this
supersoft source.}
\setlength{\tabcolsep}{6pt}
\end{deluxetable}

\clearpage
\end{landscape}

\clearpage
\tabletypesize{\footnotesize}
\LongTables
\begin{deluxetable}{lcrrrr}
\setlength{\tabcolsep}{6pt}
\tablewidth{0pt}
\tablecaption{States of Variable X-ray Sources
\label{tab:n4697x_state}}
\tablehead{
\colhead{Source}&
\colhead{Obs.$_{\rm State}$}&
\colhead{$L_{\rm State}$}&
\colhead{$H21_{\rm State}^{0}$}&
\colhead{$H31_{\rm State}^{0}$}&
\colhead{$H32_{\rm State}^{0}$}
\\
\colhead{(1)}&
\colhead{(2)}&
\colhead{(3)}&
\colhead{(4)}&
\colhead{(5)}&
\colhead{(6)}
}
\startdata
\multicolumn{6}{c}{States Determined by Luminosity Selection}\\
\phn11 & ABCD & $ 11.7\pm 1.2$ & $-$$0.32^{+0.09}_{-0.08}$ & $-$$0.94^{+0.03}_{-0.03}$ & $-$$0.88^{+0.06}_{-0.06}$\\
    &    E & $  3.2\pm 1.5$ & $-$$0.91^{+0.15}_{-0.09}$ & $-$$1.00^{+0.16}_{-0.00}$ & $-$$1.00^{+2.00}_{-0.00}$\\
\phn23 & ABCD & $ -0.2\pm 0.3$ & $+$$0.00^{+1.00}_{-1.00}$ & $+$$1.00^{+0.00}_{-1.59}$ & $+$$1.00^{+0.00}_{-2.00}$\\
    &    E & $  4.9\pm 1.8$ & $-$$1.00^{+0.09}_{-0.00}$ & $-$$1.00^{+0.08}_{-0.00}$ & $-$$1.00^{+2.00}_{-0.00}$\\
\phn25 &    A & $ 16.9\pm 2.8$ & $-$$0.94^{+0.04}_{-0.05}$ & $-$$1.00^{+0.04}_{-0.00}$ & $-$$1.00^{+1.00}_{-0.00}$\\
    & BCDE & $  4.9\pm 0.8$ & $-$$0.95^{+0.03}_{-0.04}$ & $-$$1.00^{+0.02}_{-0.00}$ & $-$$1.00^{+1.00}_{-0.00}$\\
\phn37 & ABCD & $  0.3\pm 0.4$ & $-$$0.56^{+1.54}_{-0.44}$ & $-$$0.46^{+1.44}_{-0.54}$ & $+$$0.14^{+0.86}_{-1.14}$\\
    &    E & $  8.6\pm 2.3$ & $-$$0.12^{+0.24}_{-0.21}$ & $-$$0.74^{+0.18}_{-0.17}$ & $-$$0.68^{+0.20}_{-0.21}$\\
\phn39 &    A & $ 38.9\pm 4.0$ & $+$$0.10^{+0.11}_{-0.11}$ & $-$$0.15^{+0.12}_{-0.12}$ & $-$$0.25^{+0.10}_{-0.11}$\\
    & BCDE & $ 19.0\pm 1.5$ & $-$$0.01^{+0.09}_{-0.09}$ & $-$$0.26^{+0.09}_{-0.09}$ & $-$$0.25^{+0.08}_{-0.08}$\\
\phn41 &    A & $  0.2\pm 0.8$ & $+$$1.00^{+0.00}_{-2.00}$ & $-$$1.00^{+2.00}_{-0.00}$ & $-$$1.00^{+2.00}_{-0.00}$\\
    & BCDE & $  3.4\pm 0.7$ & $-$$0.25^{+0.23}_{-0.22}$ & $-$$0.11^{+0.24}_{-0.20}$ & $+$$0.14^{+0.22}_{-0.19}$\\
\phn45 &    A & $  7.4\pm 2.0$ & $-$$0.29^{+0.21}_{-0.22}$ & $-$$0.80^{+0.17}_{-0.17}$ & $-$$0.67^{+0.28}_{-0.29}$\\
    & BCDE & $  0.1\pm 0.3$ & $+$$1.00^{+0.00}_{-0.52}$ & $+$$0.00^{+1.00}_{-1.00}$ & $-$$1.00^{+0.67}_{-0.00}$\\
\phn51 &  ABC & $  5.3\pm 1.0$ & $-$$0.33^{+0.17}_{-0.15}$ & $-$$0.76^{+0.11}_{-0.12}$ & $-$$0.57^{+0.17}_{-0.19}$\\
    &   DE & $  0.1\pm 0.6$ & $+$$0.98^{+0.02}_{-1.98}$ & $-$$1.00^{+2.00}_{-0.00}$ & $-$$1.00^{+2.00}_{-0.00}$\\
\phn67 &    A & $ -0.3\pm 0.7$ & $+$$0.00^{+1.00}_{-1.00}$ & $+$$1.00^{+0.00}_{-2.00}$ & $+$$1.00^{+0.00}_{-2.00}$\\
    & BCDE & $  8.2\pm 1.0$ & $+$$0.05^{+0.16}_{-0.14}$ & $-$$0.17^{+0.16}_{-0.15}$ & $-$$0.21^{+0.12}_{-0.13}$\\
\phn68 &   AE & $  7.8\pm 1.5$ & $-$$0.24^{+0.18}_{-0.18}$ & $-$$0.36^{+0.18}_{-0.17}$ & $-$$0.13^{+0.20}_{-0.20}$\\
    &   BC & $ 18.3\pm 2.1$ & $-$$0.35^{+0.11}_{-0.11}$ & $-$$0.42^{+0.11}_{-0.10}$ & $-$$0.08^{+0.13}_{-0.13}$\\
    &    D & $ -0.1\pm 0.9$ & $+$$1.00^{+0.00}_{-2.00}$ & $+$$1.00^{+0.00}_{-2.00}$ & $+$$0.00^{+1.00}_{-1.00}$\\
\phn77 &   AB & $  0.1\pm 0.5$ & $+$$0.96^{+0.04}_{-1.96}$ & $-$$1.00^{+2.00}_{-0.00}$ & $-$$1.00^{+2.00}_{-0.00}$\\
    &  CDE & $  8.3\pm 1.3$ & $-$$0.28^{+0.14}_{-0.13}$ & $-$$0.78^{+0.09}_{-0.09}$ & $-$$0.64^{+0.14}_{-0.13}$\\
\phn78 &    A & $ 14.9\pm 2.6$ & $-$$1.00^{+0.04}_{-0.00}$ & $-$$1.00^{+0.04}_{-0.00}$ & $+$$0.00^{+1.00}_{-1.00}$\\
    & BCDE & $  3.3\pm 0.7$ & $-$$1.00^{+0.03}_{-0.00}$ & $-$$1.00^{+0.04}_{-0.00}$ & $+$$1.00^{+0.00}_{-2.00}$\\
\phn79 & ACDE & $  1.4\pm 0.5$ & $-$$0.39^{+0.33}_{-0.26}$ & $-$$0.71^{+0.28}_{-0.26}$ & $-$$0.44^{+0.44}_{-0.48}$\\
    &    B & $  5.5\pm 1.9$ & $+$$0.19^{+0.42}_{-0.37}$ & $-$$0.04^{+0.50}_{-0.39}$ & $-$$0.22^{+0.31}_{-0.31}$\\
\phn94 &  ABC & $  0.3\pm 0.4$ & $-$$0.84^{+0.59}_{-0.16}$ & $-$$1.00^{+0.51}_{-0.00}$ & $-$$1.00^{+2.00}_{-0.00}$\\
    &   DE & $  5.3\pm 1.3$ & $-$$0.59^{+0.16}_{-0.14}$ & $-$$0.92^{+0.08}_{-0.08}$ & $-$$0.73^{+0.27}_{-0.27}$\\
\phn98 &   AE & $ 17.1\pm 1.9$ & $-$$0.37^{+0.10}_{-0.09}$ & $-$$0.67^{+0.09}_{-0.08}$ & $-$$0.40^{+0.14}_{-0.14}$\\
    &  BCD & $  8.7\pm 1.3$ & $-$$0.12^{+0.14}_{-0.14}$ & $-$$0.57^{+0.13}_{-0.12}$ & $-$$0.49^{+0.14}_{-0.13}$\\
104 & ABCD & $ -0.2\pm 0.3$ & $+$$1.00^{+0.00}_{-1.00}$ & $+$$1.00^{+0.00}_{-1.00}$ & $+$$0.95^{+0.05}_{-1.95}$\\
    &    E & $  5.2\pm 1.9$ & $-$$0.44^{+0.30}_{-0.28}$ & $-$$0.46^{+0.32}_{-0.26}$ & $-$$0.02^{+0.43}_{-0.38}$\\
110 & ABCD & $ -0.1\pm 0.4$ & $-$$1.00^{+1.00}_{-0.00}$ & $-$$1.00^{+1.00}_{-0.00}$ & $+$$0.00^{+1.00}_{-1.00}$\\
    &    E & $ 32.5\pm 4.0$ & $-$$0.99^{+0.01}_{-0.01}$ & $-$$0.99^{+0.01}_{-0.01}$ & $-$$0.26^{+1.26}_{-0.74}$\\
114 &  ABD & $ 26.9\pm 2.0$ & $-$$0.23^{+0.08}_{-0.07}$ & $-$$0.58^{+0.07}_{-0.07}$ & $-$$0.41^{+0.08}_{-0.08}$\\
    &   CE & $ 17.5\pm 2.1$ & $-$$0.19^{+0.12}_{-0.11}$ & $-$$0.64^{+0.10}_{-0.09}$ & $-$$0.52^{+0.11}_{-0.11}$\\
117 &   AC & $ 89.8\pm 4.3$ & $-$$0.06^{+0.05}_{-0.05}$ & $-$$0.41^{+0.05}_{-0.05}$ & $-$$0.36^{+0.05}_{-0.05}$\\
    &   BE & $122.6\pm 5.0$ & $-$$0.09^{+0.05}_{-0.05}$ & $-$$0.34^{+0.05}_{-0.05}$ & $-$$0.26^{+0.04}_{-0.05}$\\
    &    D & $ 40.0\pm 4.8$ & $-$$0.01^{+0.18}_{-0.17}$ & $+$$0.27^{+0.15}_{-0.14}$ & $+$$0.28^{+0.13}_{-0.11}$\\
121 &    A & $  6.5\pm 1.8$ & $-$$0.55^{+0.18}_{-0.20}$ & $-$$0.89^{+0.09}_{-0.11}$ & $-$$0.68^{+0.30}_{-0.32}$\\
    & BCDE & $  0.0\pm 0.4$ & $+$$1.00^{+0.00}_{-0.33}$ & $+$$0.00^{+1.00}_{-1.00}$ & $-$$1.00^{+0.32}_{-0.00}$\\
122 &    A & $ 11.2\pm 2.4$ & $-$$0.43^{+0.16}_{-0.16}$ & $-$$0.91^{+0.10}_{-0.09}$ & $-$$0.78^{+0.23}_{-0.22}$\\
    & BCDE & $  5.0\pm 1.0$ & $-$$0.21^{+0.19}_{-0.17}$ & $-$$0.55^{+0.18}_{-0.16}$ & $-$$0.39^{+0.20}_{-0.20}$\\
130 &  ABD & $ 23.3\pm 1.9$ & $-$$0.18^{+0.08}_{-0.08}$ & $-$$0.56^{+0.07}_{-0.07}$ & $-$$0.42^{+0.09}_{-0.09}$\\
    &   CE & $ 13.2\pm 1.9$ & $-$$0.40^{+0.12}_{-0.11}$ & $-$$0.68^{+0.11}_{-0.10}$ & $-$$0.39^{+0.16}_{-0.16}$\\
131 & ABCD & $  4.0\pm 0.8$ & $+$$0.10^{+0.24}_{-0.21}$ & $-$$0.18^{+0.26}_{-0.23}$ & $-$$0.28^{+0.19}_{-0.19}$\\
    &    E & $ 11.3\pm 2.6$ & $+$$0.14^{+0.28}_{-0.24}$ & $-$$0.06^{+0.31}_{-0.26}$ & $-$$0.20^{+0.21}_{-0.20}$\\
134 &  ACE & $161.9\pm 4.7$ & $+$$0.67^{+0.03}_{-0.03}$ & $+$$0.64^{+0.04}_{-0.04}$ & $-$$0.07^{+0.03}_{-0.03}$\\
    &   BD & $188.8\pm 6.8$ & $+$$0.54^{+0.05}_{-0.05}$ & $+$$0.55^{+0.05}_{-0.05}$ & $+$$0.02^{+0.04}_{-0.04}$\\
144 &  ABD & $  2.9\pm 0.9$ & $-$$0.03^{+0.35}_{-0.30}$ & $-$$0.26^{+0.39}_{-0.32}$ & $-$$0.23^{+0.32}_{-0.31}$\\
    &   CE & $  8.4\pm 1.6$ & $+$$0.06^{+0.23}_{-0.22}$ & $-$$0.12^{+0.26}_{-0.22}$ & $-$$0.18^{+0.19}_{-0.17}$\\
147 &  BCE & $  4.1\pm 1.0$ & $+$$0.07^{+0.28}_{-0.23}$ & $-$$0.52^{+0.30}_{-0.24}$ & $-$$0.57^{+0.21}_{-0.22}$\\
    &    D & $ 13.7\pm 3.2$ & $-$$0.24^{+0.22}_{-0.18}$ & $-$$0.76^{+0.17}_{-0.14}$ & $-$$0.64^{+0.21}_{-0.21}$\\
155 &   BC & $ 92.3\pm 4.8$ & $-$$0.04^{+0.06}_{-0.06}$ & $-$$0.33^{+0.06}_{-0.06}$ & $-$$0.29^{+0.05}_{-0.05}$\\
    &    D & $121.6\pm 8.4$ & $-$$0.00^{+0.08}_{-0.08}$ & $-$$0.26^{+0.08}_{-0.08}$ & $-$$0.26^{+0.07}_{-0.07}$\\
156 &    B & $117.1\pm 7.5$ & $-$$0.74^{+0.03}_{-0.03}$ & $-$$0.99^{+0.01}_{-0.01}$ & $-$$0.95^{+0.06}_{-0.05}$\\
    &   CD & $ 74.5\pm 4.6$ & $-$$0.84^{+0.03}_{-0.02}$ & $-$$1.00^{+0.01}_{-0.00}$ & $-$$1.00^{+0.06}_{-0.00}$\\
\newpage
\multicolumn{6}{c}{States Determined by $H_{21}^0$ Selection}\\
\phn11 &  ABE & $  8.7\pm 1.2$ & $-$$0.56^{+0.09}_{-0.09}$ & $-$$0.95^{+0.03}_{-0.04}$ & $-$$0.84^{+0.09}_{-0.13}$\\
    &   CD & $ 12.2\pm 1.9$ & $-$$0.14^{+0.14}_{-0.13}$ & $-$$0.93^{+0.06}_{-0.07}$ & $-$$0.91^{+0.08}_{-0.09}$\\
\phn63 &  ACD & $ 26.1\pm 2.0$ & $+$$0.09^{+0.09}_{-0.09}$ & $-$$0.27^{+0.10}_{-0.09}$ & $-$$0.35^{+0.08}_{-0.08}$\\
    &   BE & $ 30.0\pm 2.6$ & $-$$0.28^{+0.09}_{-0.09}$ & $-$$0.50^{+0.08}_{-0.08}$ & $-$$0.26^{+0.10}_{-0.10}$\\
103 &   AC & $ 35.0\pm 2.8$ & $+$$0.14^{+0.09}_{-0.09}$ & $-$$0.36^{+0.10}_{-0.09}$ & $-$$0.47^{+0.08}_{-0.08}$\\
    &  BDE & $ 33.6\pm 2.3$ & $-$$0.33^{+0.07}_{-0.07}$ & $-$$0.47^{+0.07}_{-0.06}$ & $-$$0.18^{+0.08}_{-0.08}$\\
134 &   AE & $158.3\pm 5.5$ & $+$$0.73^{+0.04}_{-0.04}$ & $+$$0.72^{+0.04}_{-0.04}$ & $-$$0.02^{+0.03}_{-0.03}$\\
    &  BCD & $182.8\pm 5.4$ & $+$$0.55^{+0.04}_{-0.04}$ & $+$$0.52^{+0.04}_{-0.04}$ & $-$$0.04^{+0.03}_{-0.03}$\\
140 &   AD & $  3.4\pm 1.4$ & $+$$0.83^{+0.17}_{-0.45}$ & $+$$0.75^{+0.25}_{-0.64}$ & $-$$0.23^{+0.34}_{-0.34}$\\
    &  BCE & $  1.4\pm 0.7$ & $-$$0.74^{+0.24}_{-0.24}$ & $-$$0.91^{+0.14}_{-0.09}$ & $-$$0.54^{+0.82}_{-0.46}$\\
152 &   AC & $  1.2\pm 1.1$ & $+$$1.00^{+0.00}_{-0.00}$ & $+$$1.00^{+0.00}_{-1.00}$ & $-$$0.74^{+0.46}_{-0.26}$\\
    &   BD & $  3.6\pm 1.4$ & $-$$0.28^{+0.34}_{-0.28}$ & $-$$0.55^{+0.36}_{-0.32}$ & $-$$0.32^{+0.41}_{-0.47}$\\
156 &    B & $117.1\pm 7.5$ & $-$$0.74^{+0.04}_{-0.03}$ & $-$$0.99^{+0.01}_{-0.01}$ & $-$$0.95^{+0.06}_{-0.05}$\\
    &   CD & $ 74.5\pm 4.6$ & $-$$0.84^{+0.03}_{-0.02}$ & $-$$1.00^{+0.01}_{-0.00}$ & $-$$1.00^{+0.06}_{-0.00}$\\
\\
\multicolumn{6}{c}{States Determined by $H_{31}^0$ Selection}\\
\phn10 & ABCE & $ 19.1\pm 1.5$ & $-$$0.08^{+0.09}_{-0.09}$ & $-$$0.28^{+0.09}_{-0.08}$ & $-$$0.20^{+0.08}_{-0.08}$\\
    &    D & $ 14.7\pm 3.1$ & $-$$0.23^{+0.18}_{-0.17}$ & $-$$0.87^{+0.09}_{-0.11}$ & $-$$0.80^{+0.12}_{-0.16}$\\
\phn48 & ACDE & $  6.4\pm 1.0$ & $-$$0.02^{+0.17}_{-0.15}$ & $-$$0.32^{+0.18}_{-0.15}$ & $-$$0.31^{+0.15}_{-0.14}$\\
    &    B & $  4.2\pm 1.8$ & $-$$0.26^{+0.31}_{-0.29}$ & $-$$1.00^{+0.16}_{-0.00}$ & $-$$1.00^{+0.26}_{-0.00}$\\
\phn58 & ABCE & $ 59.3\pm 2.4$ & $-$$0.10^{+0.05}_{-0.05}$ & $-$$0.34^{+0.05}_{-0.05}$ & $-$$0.24^{+0.05}_{-0.05}$\\
    &    D & $ 62.2\pm 5.8$ & $+$$0.19^{+0.12}_{-0.12}$ & $+$$0.05^{+0.13}_{-0.12}$ & $-$$0.15^{+0.09}_{-0.09}$\\
\phn66 & ABCE & $  9.5\pm 1.1$ & $-$$0.09^{+0.12}_{-0.12}$ & $-$$0.35^{+0.12}_{-0.12}$ & $-$$0.26^{+0.11}_{-0.12}$\\
    &    D & $ 11.4\pm 2.8$ & $+$$0.16^{+0.61}_{-0.38}$ & $+$$0.62^{+0.30}_{-0.20}$ & $+$$0.50^{+0.21}_{-0.19}$\\
102 & ABCE & $  1.6\pm 0.5$ & $+$$0.15^{+0.39}_{-0.30}$ & $-$$0.56^{+0.48}_{-0.38}$ & $-$$0.65^{+0.27}_{-0.30}$\\
    &    D & $  1.3\pm 1.6$ & $+$$0.00^{+1.00}_{-1.00}$ & $+$$1.00^{+0.00}_{-0.33}$ & $+$$1.00^{+0.00}_{-0.57}$\\
117 & ABCE & $106.1\pm 3.3$ & $-$$0.08^{+0.04}_{-0.03}$ & $-$$0.37^{+0.04}_{-0.03}$ & $-$$0.30^{+0.03}_{-0.03}$\\
    &    D & $ 40.0\pm 4.8$ & $-$$0.01^{+0.18}_{-0.17}$ & $+$$0.27^{+0.15}_{-0.14}$ & $+$$0.28^{+0.12}_{-0.11}$\\
134 &  ABE & $169.6\pm 4.7$ & $+$$0.68^{+0.03}_{-0.03}$ & $+$$0.69^{+0.03}_{-0.03}$ & $+$$0.01^{+0.03}_{-0.03}$\\
    &   CD & $175.8\pm 6.7$ & $+$$0.53^{+0.05}_{-0.05}$ & $+$$0.44^{+0.06}_{-0.05}$ & $-$$0.11^{+0.04}_{-0.04}$\\
\\
\multicolumn{6}{c}{States Determined by $H_{32}^0$ Selection}\\
\phn10 & ABCE & $ 19.1\pm 1.5$ & $-$$0.08^{+0.09}_{-0.09}$ & $-$$0.28^{+0.09}_{-0.09}$ & $-$$0.20^{+0.08}_{-0.08}$\\
    &    D & $ 14.7\pm 3.1$ & $-$$0.23^{+0.18}_{-0.18}$ & $-$$0.87^{+0.09}_{-0.11}$ & $-$$0.80^{+0.13}_{-0.16}$\\
\phn66 &   AD & $ 10.1\pm 1.6$ & $-$$0.24^{+0.19}_{-0.19}$ & $+$$0.10^{+0.17}_{-0.17}$ & $+$$0.33^{+0.16}_{-0.17}$\\
    &  BCE & $  9.7\pm 1.3$ & $-$$0.00^{+0.15}_{-0.14}$ & $-$$0.35^{+0.15}_{-0.15}$ & $-$$0.35^{+0.12}_{-0.13}$\\
\phn81 &  ABD & $  1.4\pm 0.7$ & $+$$0.21^{+0.68}_{-0.41}$ & $-$$0.56^{+0.88}_{-0.44}$ & $-$$0.69^{+0.32}_{-0.31}$\\
    &   CE & $  1.6\pm 0.9$ & $+$$1.00^{+0.00}_{-1.27}$ & $+$$1.00^{+0.00}_{-0.15}$ & $+$$0.75^{+0.25}_{-0.21}$\\
102 &  ACE & $  1.7\pm 0.7$ & $+$$0.14^{+0.38}_{-0.32}$ & $-$$0.84^{+0.43}_{-0.16}$ & $-$$0.88^{+0.27}_{-0.12}$\\
    &   BD & $  1.2\pm 0.9$ & $+$$1.00^{+0.00}_{-1.41}$ & $+$$1.00^{+0.00}_{-0.34}$ & $+$$0.77^{+0.23}_{-0.34}$\\
117 & ABCE & $106.1\pm 3.3$ & $-$$0.08^{+0.04}_{-0.03}$ & $-$$0.37^{+0.04}_{-0.03}$ & $-$$0.30^{+0.03}_{-0.03}$\\
    &    D & $ 40.0\pm 4.8$ & $-$$0.01^{+0.18}_{-0.17}$ & $+$$0.27^{+0.15}_{-0.14}$ & $+$$0.28^{+0.12}_{-0.11}$\\
134 & ABDE & $171.9\pm 4.3$ & $+$$0.63^{+0.03}_{-0.03}$ & $+$$0.63^{+0.03}_{-0.03}$ & $-$$0.00^{+0.03}_{-0.02}$\\
    &    C & $171.0\pm 9.3$ & $+$$0.57^{+0.07}_{-0.06}$ & $+$$0.44^{+0.08}_{-0.08}$ & $-$$0.17^{+0.05}_{-0.05}$\\
151 &    A & $ 11.5\pm 2.4$ & $-$$0.10^{+0.19}_{-0.19}$ & $-$$0.82^{+0.16}_{-0.16}$ & $-$$0.78^{+0.19}_{-0.18}$\\
    &    E & $ 11.6\pm 5.4$ & $-$$0.76^{+0.29}_{-0.24}$ & $-$$0.34^{+0.41}_{-0.31}$ & $+$$0.56^{+0.44}_{-0.42}$\\
\enddata
\setlength{\tabcolsep}{6pt}
\end{deluxetable}


\end{document}